\documentclass[nofootinbib,aps,11pt]{revtex4-1}
\usepackage{graphicx}
\usepackage{hyperref}
\usepackage{cancel}
\usepackage{amssymb}
\usepackage{textcomp}
\usepackage{amsmath}
\usepackage{bm}
\usepackage{times}
\usepackage{epsfig}
\usepackage{color}
\usepackage{slashed}
\newcommand{\hc}{{\rm h.c.}}

\newcommand{\MeV}{{\rm MeV}}
\newcommand{\GeV}{{\rm GeV}}
\newcommand{\TeV}{{\rm TeV}}
\newcommand{\fb}{{\rm fb}}
\newcommand{\BR}{{\rm BR}}
\newcommand{\calO}{{\cal O}}
\def\lsim{\mathrel{\rlap{\lower4pt\hbox{\hskip1pt$\sim$}}
    \raise1pt\hbox{$<$}}}    % less than or approx. symbol

\begin{document}
\title{\LARGE  Gauged $U(1)_{L_\mu-L_\tau}$ Scotogenic Model in light of $R_{K^{(*)}}$ Anomaly and AMS-02 Positron Excess}
\bigskip

\author{Zhi-Long Han$^1$}
\email{sps\_hanzl@ujn.edu.cn}
\author{Ran Ding$^2$}
\email{dingran@mail.nankai.edu.cn}
\author{Su-Jie Lin$^3$}
\email{linsj@ihep.ac.cn}
\author{Bin Zhu$^4$}
\email{zhubin@mail.nankai.edu.cn}

\affiliation{
$^1$School of Physics and Technology, University of Jinan, Jinan, Shandong 250022, China
\\
$^2$Center for High Energy Physics, Peking University, Beijing 100871, China
\\
$^3$Key Laboratory of Particle Astrophysics, Institute of High
Energy Physics,
Chinese Academy of Sciences, Beijing 100049, China
\\
$^4$Department of Physics, Yantai University, Yantai 264005, P. R. China}

\date{\today}

\begin{abstract}
  We study the gauged $U(1)_{L_\mu-L_\tau}$ scotogenic model with emphasis on latest measurement of LHCb $R_{K^{(*)}}$ anomaly and AMS-02 positron excess. In this model, neutrino masses are induced at one-loop level with $Z_2$-odd particles, i.e., right-handed neutrinos $N_\ell(\ell=e,\mu,\tau)$ and inert scalar doublet $\eta$ inside the loop. Meanwhile, the gauged $U(1)_{L_\mu-L_\tau}$ symmetry is broken spontaneously by the scalar singlet $S$, resulting to the massive gauge boson $Z'$. Provided certain couplings to quarks induced by heavy vector-like quarks, the gauge boson $Z'$ would contribute to the transition $b\to s \mu^+\mu^-$, hence explain the $R_{K^{(*)}}$ anomaly. As for the Majorana fermion DM $N$, the gauge boson $Z'$ and the singlet Higgs $H_0$ will generate various annihilation channels, among which the $NN\to Z'Z'$ and $NN\to Z'H_0(\to Z'Z')$ channel could be used to interpret the AMS-02 positron excess. We give a comprehensive analysis on model parameter space with consider various current constraints.
  The combined analysis shows that the $R_{K^{(*)}}$ anomaly and AMS-02 positron excess can be explained simultaneously.
\end{abstract}

\maketitle

%%%%%%%%%%%%%%%%%%%%%%%
\section{Introduction}
%%%%%%%%%%%%%%%%%%%%%%%

Tiny neutrino mass and non-baryonic dark matter (DM) are the two missing pieces of standard model (SM). An appealing pathway to link them together is the scotogenic model \cite{Krauss:2002px,Ma:2006km}, which realizes tiny neutrino mass via radiative process~\cite{Zee:1980ai} with DM running in the loop. Along this idea, various possibilities~\cite{scotogenic} have been proposed~\footnote{See Ref.~\cite{Cai:2017jrq} for a recent review and more references therein.}. However, in the minimal version of scotogenic model, the parameter space of fermionic DM required by relic abundance is tightly constrained by lepton flavor violation (LFV) processes~\cite{Kubo:2006yx}. Such tension motivates the suggestions that extend the original model with a new $U(1)$ gauge group. The contradiction is then relaxed due to new available annihilation channels via exchanging of gauge or Higgs boson associated with the $U(1)$ group. Particularly, comparing with gauged $U(1)_{B-L}$ model~\cite{BL}, gauged $U(1)_{L_\mu-L_\tau}$ model ~\cite{Baek:2008nz,Heeck:2011wj,Das:2013jca,Baek:2015mna,Baek:2015fea,
Biswas:2016yan,Biswas:2016yjr,Lee:2017ekw,Asai:2017ryy,Dev:2017fdz,Biswas:2017ait,Nomura:2018vfz,
Nomura:2018cle,Banerjee:2018eaf,Kamada:2018zxi,Foldenauer:2018zrz}) has less stringent constraints due to the fact that corresponding gauge boson $Z^\prime$ does not couple to SM quarks and electron directly. It is worthy to note that a light $Z^\prime\sim\mathcal{O}(100)~\MeV$ with gauge coupling $g^\prime\sim10^{-3}$ is suitable to interpret the muon $g-2$ anomaly \cite{Baek:2001kca,Ma:2001md}.

Except the evidences from neutrino mass and DM, the hint from flavor physics may also call for the physics beyond standard model (BSM). Recently, a tentative evidence indicates lepton flavor universality (LFU) violation has been reported by LHCb Collaboration in the semi-leptonic decays of the $B$ meson. The latest result gives~\cite{Aaij:2019wad}
\begin{equation}
R_K =\frac{\text{Br}(B\to K \mu^+\mu^-)}{\text{Br}(B\to K e^+e^-)}
=0.846^{+0.060+0.016}_{-0.054-0.014}\quad \text{for}\quad 1.1<q^2<6~\GeV^2,
\end{equation}
where $q^2$ is squared momentum of the leptonic system. This result presents $2.5\sigma$ deviation with respect to the SM prediction~\cite{Hiller:2003js}.
In addition, observations of a tension in angular observable, such as $P_5^\prime$ in the decay $B\to K^\ast\mu^+\mu^-$ and angular distribution in the decay $B_s^0\to \phi \mu^+\mu^-$, have been announced by LHCb~\cite{Aaij:2015oid} and Belle~\cite{Wehle:2016yoi} as well.
Since the decay process $b\to s\ell\ell(\ell=e,\mu)$ is involved in the aforementioned $B$ anomalies, a new physics contribution to the corresponding Wilson coefficient is able to explain such anomaly~\cite{RK}. Especially, the LFU violation and angular anomaly in $b\to s\mu^+\mu^-$ indicate that the new physics particles may prefer couple to muon rather than to electron, which is actually an intrinsic feature of $U(1)_{L_\mu-L_\tau}$ gauge boson $Z^\prime$~\cite{RKZ}.

In addition, the latest direct detection experiments, such as LUX~\cite{Akerib:2016vxi}, XENON1T~\cite{Aprile:2017iyp} and PandaX-II~\cite{Cui:2017nnn} remains  for DM signal. Since these experiments based on DM-hadron interaction, the null results of DM direct detection signal may suggest that the DM-hadron interaction is at least suppressed or even better vanishing. On the other hand, the indirect detection experiments, such as PAMELA~\cite{Adriani:2013uda}, Fermi-LAT~\cite{FermiLAT:2011ab}, and AMS-02~\cite{Aguilar:2013qda,Accardo:2014lma,Aguilar:2014mma}, have reported a significance positron fraction excess in the cosmic-ray, while no obvious antiproton excess is observed~\cite{Aguilar:2016kjl}. This can be interpreted by DM dominantly annihilating into leptonic final states~\cite{Cirelli:2008pk,Feng:2017tnz}. Therefore, the current direct and indirect detection experiments seem to advocate `leptophilic DM'~\cite{Fox:2008kb}. And $U(1)_{L_\mu-L_\tau}$ DM is clearly a good choice \cite{Baek:2008nz}.

With holding such benefits, we thus consider the phenomenology of the gauged $U(1)_{L_\mu-L_\tau}$ scotogenic model \cite{Baek:2015mna} in this paper, which has been studied in Refs.~\cite{Baek:2015mna,Baek:2015fea} with emphasizing neutrino mass and dark matter properties for light $Z^\prime$. According to Ref.~\cite{Baek:2015fea}, $50~\MeV\lesssim M_{Z'}\lesssim400~\MeV$ with $3\times10^{-4}\lesssim g'\lesssim10^{-3}$ is required to explain $(g-2)_\mu$ and DM relic density. Lately, searches for $Z'$ in the $e^+e^-\to \mu^+\mu^- Z', Z'\to \mu^+\mu^-$ channel by BABAR has rejected $M_{Z'}>212~\MeV$ with $g'\gtrsim7\times10^{-4}$ \cite{TheBABAR:2016rlg}. Although not fully excluded at current, the future experiments, such as Belle-II and SPS, are able to exclude the whole low $Z'$ mass region favored by $(g-2)_\mu$ and DM \cite{Gninenko:2014pea,Kaneta:2016uyt,Araki:2017wyg,
Chen:2017cic}. Therefore, we move on to the high mass region $M_{Z'}\gtrsim10~\GeV$ and specially focus on $R_{K^{(*)}}$ anomaly and AMS-02 positron excess, which has not been discussed in previous studies \cite{Baek:2015mna,Baek:2015fea}.

This paper is organized as following. In Sec.~\ref{sec:model}, we briefly review the gauged $U(1)_{L_\mu-L_\tau}$ scotogenic model. Interpretation of the $R_{K^{(*)}}$ anomaly is discussed in Sec.~\ref{sec:RK}, and relative constraints on $Z'$ in the high mass region are summarised in Sec.~\ref{sec:CT}. A detail scanning on the DM parameter space under constraints from relic density and direct detection are performed in Sec.~\ref{sec:DM}. In Sec.~\ref{sec:AMS}, we provide benchmarks to fit positron excess by using latest AMS-02 measurement and discuss relative constraints imposed by indirect detections. Finally, Sec.~\ref{sec:CL} is devoted to conclusion.

%%%%%%%%%%%%%%%%%%%%%%%
\section{The Model}\label{sec:model}
%%%%%%%%%%%%%%%%%%%%%%%

\subsection{Model Setup}
The gauged $U(1)_{L\mu-L_\tau}$ scotogenic model was proposed in Ref.~\cite{Baek:2015mna}.
Comparing with the original scotogenic model~\cite{Ma:2006km}, this model further  introduces one additional singlet scalar $S$ charged $+1$ under $U(1)_{L_\mu-L_\tau}$ to break this symmetry spontaneously. The particle content and the charge assignment under $SU(2)_L \times U(1)_Y \times U(1)_{L_\mu-L_\tau} \times Z_2$ are shown in table~\ref{tab:chargeMa}. Here, the right-handed neutrinos $N_{\ell}(\ell=e,\mu,\tau)$ and inert scalar doublet $\eta$ have $Z_2$ odd charge, thus the lightest of them can be a dark matter candidate. In this paper, we consider fermion DM. Neutrino masses are generated at one-loop level via $N_\ell$ and $\eta$ propagating in the loop as the original scotogenic model.

\begin{center}
\begin{table}[!htbp]
\begin{tabular}{|c|c|c|c|c|c|c|c|c|c|c|c|c|}
\hline
          & $L_L^e$ & $L_L^\mu$ & $L_L^\tau$ & $e_R^c$ & $\mu_R^c$ & $\tau_R^c$ &$N_e^c$ &$N_\mu^c$ &$N_\tau^c$  &~$\Phi$~  & ~$\eta$& ~$S$~   \\\hline
$SU(2)_L$ & \multicolumn{3}{c|}{$\bm{2}$} & \multicolumn{3}{c|}{$\bm{1}$}&\multicolumn{3}{c|}{ $\bm{1}$} & $\bm{2}$&  $\bm{2}$ & $\bm{1}$  \\\hline
$U(1)_Y$   & \multicolumn{3}{c|}{$-1/2$}   & \multicolumn{3}{c|}{$1$} & \multicolumn{3}{c|}{$0$} & $+1/2$ & $+1/2$ & $0$  \\\hline
$U(1)_{L_\mu - L_\tau}$  & $0$ & $+1$ & $-1$ & $0$ & $-1$ & $+1$ & $0$ & $-1$ & $+1$ & $0$ & $0$ & $+1$ \\ \hline
$Z_2$  & \multicolumn{3}{c|}{$+$} & \multicolumn{3}{c|}{$+$} & \multicolumn{3}{c|}{$-$} & $+$& $-$& $+$ \\\hline
\end{tabular}
\caption{The particle content and the charge assignment under $SU(2)_L \times U(1)_Y \times U(1)_{L_\mu-L_\tau}\! \times Z_2$.}
\label{tab:chargeMa}
\end{table}
\end{center}

The scalar potential involving scalar doublet $\Phi$, scalar singlet $S$ and inert scalar doublet $\eta$ is ~\cite{Baek:2015mna}
\begin{eqnarray}
V&=&  + \mu_\Phi^2 |\Phi|^2 + \mu_\eta^2 |\eta|^2 + \mu_S^2 |S|^2 + \frac{1}{2} \lambda_1 |\Phi|^4  + \frac{1}{2} \lambda_2 |\eta|^4   \\ \nonumber
&& + \lambda_3 |\Phi|^2 |\eta|^2 + \lambda_4 |\Phi^\dagger \eta|^2 +\frac{1}{2} \lambda_5 \left[ (\Phi^\dagger \eta)^2 + {\rm h.c.} \right]
 \\ \nonumber
 &&  + \lambda_6 |S|^4 + \lambda_7 |S|^2 |\Phi|^2 + \lambda_8 |S|^2 |\eta|^2.
\end{eqnarray}
After SSB, the scalar fields $\Phi,~\eta,~S$ are denoted as
\begin{align}
\Phi = \left(\begin{array}{cc}
G^+ \\
\frac{1}{\sqrt{2}}(v + \varphi + iG^0)
\end{array}\right),\quad
\eta = \left(\begin{array}{cc}
\eta^+ \\
\frac{1}{\sqrt{2}}(\eta_H + i\eta_A)
\end{array}\right),\quad
S= \frac{1}{\sqrt{2}}(v_S + s + iG_S),  \label{component}
\end{align}
where $v=246$ GeV, $v_S$ is the VEV of $S$ which breaks the $U(1)_{L_\mu-L_\tau}$ symmetry. $G^\pm$, $G^0$ and $G_S$ are corresponding Nambu-Goldstone bosons, which are respectively absorbed by the longitudinal component of the $W^\pm$, $Z$ and $Z^\prime$ gauge bosons. Therefore the VEV $v_S$ provides a mass of $U(1)_{L_\mu-L_\tau}$ gauge boson $Z'$ as $M_{Z'}=g'v_S$. Due to the unbroken $Z_2$ symmetry, the components of inert doublet field  $\eta$ ($\eta^{\pm},~\eta_H,~\eta_A$) do not mix with other scalar fields and their squared masses are simply given as
\begin{align}
M_{\eta^\pm}^2 &= \mu_\eta^2 +\frac{v^2}{2}\lambda_3 +\frac{v_S^2}{2}\lambda_8, \\
M_{\eta_A}^2 &= \mu_\eta^2 +\frac{v^2}{2}(\lambda_3+\lambda_4-\lambda_5) +\frac{v_S^2}{2}\lambda_8, \\
M_{\eta_H}^2 &= \mu_\eta^2 +\frac{v^2}{2}(\lambda_3+\lambda_4+\lambda_5) +\frac{v_S^2}{2}\lambda_8.
\end{align}
While for the $Z_2$-even sector, two CP-even scalar components $\varphi$ and $s$ mix with each other. And in the gauge eigenstates $(\varphi,~s)$, their mass matrix $\mathcal{M}_H^2$ is written as
\begin{align}
\mathcal{M}_H^2 =
\begin{pmatrix}
v^2\lambda_1 & vv_S\lambda_7 \\
vv_S\lambda_7 & 2v_S^2\lambda_6
\end{pmatrix}. \label{mass_mat}
\end{align}
$\mathcal{M}_H^2$ can be diagonalized to give the mass eigenstates $h$ and $H_0$. The corresponding squared mass eigenvalues are
\begin{align}
M^2_{h,~H_0} =
\frac{1}{2} \left[ (\mathcal{M}_H^2)_{11} + (\mathcal{M}_H^2)_{22}
\pm \sqrt{\left((\mathcal{M}_H^2)_{11} - (\mathcal{M}_H^2)_{22}\right)^2 + 4 \left|\mathcal{M}_H^2)_{12}\right|^2 }\right]~,
\label{eq:eigenmass}
\end{align}
The mass and gauge eigenstates are related by
\begin{align}
\left\{\begin{array}{ll}h &= \varphi\cos{\alpha}+s\sin{\alpha}~,\\
H_0 &= -\varphi\sin{\alpha}+s\cos{\alpha}~,\end{array}\right.
\label{eq:mixing}
\end{align}
with the mixing angle given by
\begin{align}
\tan{2\alpha}=\frac{2 (\mathcal{M}_H^2)_{12}}{(\mathcal{M}_H^2)_{11} - (\mathcal{M}_H^2)_{22}}~.\quad
\label{eq:mixangle}
\end{align}
Here we assume $h$ as the SM-like Higgs boson with the mass of 125 GeV.
Thus, $H_0$ corresponds to an additional singlet-like Higgs boson. In the following, we mainly consider $\alpha\lesssim0.1$ to avoid various constrains \cite{Robens:2016xkb}.
Finally, the bounded from below condition requires:
\begin{align}
&\lambda_1,~\lambda_2,~\lambda_6 > 0, \\
&\lambda_7 + \frac{1}{\sqrt{2}}\sqrt{\lambda_1\lambda_6} > 0, \quad
\lambda_8 + \frac{1}{\sqrt{2}}\sqrt{\lambda_2\lambda_6} > 0, \\
&\lambda_3 + \frac{1}{2}\sqrt{\lambda_1\lambda_6}+\min(0,\lambda_4\pm\lambda_5) > 0.
\end{align}

At tree level, there is no mixing between $Z$ and $Z'$. But at one-loop level, $Z$ and $Z'$ would mix via the exchange of $\mu,\nu_\mu,\tau,\nu_\tau$ with the loop factor estimated as \cite{Patra:2016shz}
\begin{equation}
\Pi^{\mu\nu}(q^2)=-(q^2g^{\mu\nu}-q^\mu q^\nu)\frac{1}{3}\frac{1}{16\pi^2}
\left(g'\frac{C_V g}{2\cos\theta_W}\right)\log\left(\frac{M^2_\mu}{M^2_\tau}\right),
\end{equation}
where $C_V=-1+\sin^2\theta_W$ and $\theta_W$ is the Weinberg angle. The resulting mixing angle between $Z$ and $Z'$ thus is
\begin{equation}
\tan2\theta_Z=\frac{2\Pi^{\mu\nu}g_{\mu\nu}}{M^2_{Z'}-M^2_Z}.
\end{equation}
To satisfy the precise measurement of SM $Z$-boson mass \cite{Agashe:2014kda}, one needs $\tan\theta_Z<10^{-2}$\cite{Patra:2016shz}.

\subsection{Neutrino Mass}

The relevant mass terms and Yukawa interactions are flavor dependent
\begin{eqnarray}
-\mathcal{L}_Y &=&  \frac{1}{2}M_{ee} \overline{N^c_e} N_e +\frac{1}{2}M_{\mu\tau}( \overline{N^c_\mu} N_\tau + \overline{N^c_\tau} N_\mu)\\ \nonumber
&&+ h_{e\mu} (\overline{N^c_e} N_\mu + \overline{N^c_\mu} N_e) S^* + h_{e\tau} (\overline{N^c_e} N_\tau + \overline{N^c_\tau} N_e) S \\ \nonumber
&& + f_e \overline{L_L^e}  \tilde{\eta} N_e + f_\mu \overline{L_L^\mu} \tilde{\eta} N_\mu + f_\tau \overline{L_L^\tau} \tilde{\eta} N_\tau \\ \nonumber
&&+y_e \overline{L_L^e} \Phi e_R + y_\mu \overline{L_L^\mu} \Phi \mu_R + y_\tau \overline{L_L^\tau} \Phi \tau_R+\hc,
\end{eqnarray}
where $\tilde{\eta}=(i\sigma_2) \eta^*$. Hence, after $\Phi(S)$ develops VEV $v(v_S)$, mass matrix of charged lepton $\ell$ and right-handed neutrino $N_\ell$ can be written as
\begin{align}
\mathcal{M}_\ell=\frac{v}{\sqrt{2}}\text{diag}(y_e,y_\mu,y_\tau),\quad
\mathcal{M}_N=\left(
\begin{array}{ccc}
M_{ee} & \frac{v_S}{\sqrt{2}} h_{e\mu} & \frac{v_S}{\sqrt{2}} h_{e\tau}\\
\frac{v_S}{\sqrt{2}} h_{e\mu} & 0 & M_{\mu\tau}e^{i\theta_R} \\
\frac{v_S}{\sqrt{2}} h_{e\tau} & M_{\mu\tau}e^{i\theta_R} & 0
\end{array}
\right)
\end{align}
With appropriate phase redefinition, all the parameters can be made real, and $\theta_R$ is the CP-violating phase. The symmetric matrix $\mathcal{M}_N$ can be diagonalized by an unitary matrix $V$ as
\begin{equation}
V^T \mathcal{M}_N V= \text{diag}(M_1,M_2,M_3),
\end{equation}
where the lightest one is regard as DM candidate and denoted as $N$ for simplicity in the following discussion.

The neutrino mass is generated at one-loop level via exchanging $\eta$ and $N_\ell$ in the loop. Provided the inert doublet scalar much heavier than the right handed neutrinos, the resulting neutrino mass matrix is approximately given by
\begin{equation}
(\mathcal{M}_\nu)_{ij}\simeq \frac{\lambda_5 v^2}{8\pi M_0^2} f_i (\mathcal{M}_N)_{ij} f_j,
\end{equation}
where $M_0^2=(M^2_{\eta_H}+M^2_{\eta_A})/2$. Similar as $\mathcal{M}_N$, the structure of $\mathcal{M}_\nu$ corresponds to ``Pattern C" of two-zero texture in Ref.~\cite{Fritzsch:2011qv}. Therefore, only the inverted neutrino mass hierarchy can fit the neutrino oscillation data \cite{Baek:2015mna}. Due to the two-zero texture, the nine neutrino parameters are determined by five input parameters.  Briefly speaking, the heavy right-handed neutrino mass matrix is determined by $M_{ee}$, $v_S$, $h_{e\mu}$, $h_{e\tau}$ and $M_{\mu\tau}(\theta_R)$. Then for a given $\mathcal{M}_N$, the neutrino mass and mixing parameters can be acquired by tuning free parameters $\lambda_5$, $M_0$, $f_e$, $f_\mu$ and $f_\tau$, as long as the following condition is satisfied \cite{Baek:2015fea}
\begin{equation}\label{eq:R}
R=\frac{(\mathcal{M}_\nu)_{12}(\mathcal{M}_\nu)_{13}}
{(\mathcal{M}_\nu)_{11}(\mathcal{M}_\nu)_{23}}
= \frac{h_{e\mu} h_{e\tau }v_S^2e^{-i\theta_R}}{M_{ee} M_{\mu\tau}}\simeq0.46\times e^{3.1i}.
\end{equation}

\section{ $R_{K^{(*)}}$ Anomaly}\label{sec:RK}

In order to account for the $R_K$ and $R_{K^*}$ anomalies, a flavor changing coupling $Z^\prime bs$ is necessary, which is however absent in the original model due to the fact that quarks do not carry $U(1)_{\mu-\tau}$ charges. As a complement, we follow Ref.~\cite{Altmannshofer:2014cfa} to extend the original model by introducing a set of heavy vector-like quarks $Q_L \equiv(U_L,~D_L)$, $U_R^c$, $D_R^c$ and their chiral partners $\tilde Q_{R}\equiv(\tilde U_R,~\tilde D_R)$, $\tilde U_{L}^c$, $\tilde D_{L}^c$, and whose charge assignment under $SU(2)_L \times U(1)_Y \times U(1)_{\mu-\tau} \times Z_2$ are listed in table~\ref{tab:chargevq}.

\begin{center}
\begin{table}[!htbp]
\begin{tabular}{|c|c|c|c|c|c|c|}
\hline
          & $Q_{L}$ & $U_R^c$ & $D_R^c$ & $\tilde Q_{R}$ & $\tilde U_{L}^c$ & $\tilde D_{L}^c$ \\\hline
$SU(3)_C$ & $\bm{3}$ & \multicolumn{2}{c|}{$\bm{\bar{3}}$} & $\bm{3}$ & \multicolumn{2}{c|}{$\bm{\bar{3}}$}\\\hline
$SU(2)_L$ & $\bm{2}$ & \multicolumn{2}{c|}{$\bm{1}$} & $\bm{2}$ & \multicolumn{2}{c|}{$\bm{1}$} \\\hline
$U(1)_Y$  & $1/6$ & $-2/3$ & $1/3$ & $1/6$ & $-2/3$ & $1/3$  \\\hline
$U(1)_{\mu - \tau}$ & $+1$ & \multicolumn{2}{c|}{$-1$} & $+1$ & \multicolumn{2}{c|}{$-1$}  \\ \hline
$Z_2$ & \multicolumn{6}{c|}{$+$}  \\\hline
\end{tabular}
\caption{The charge assignment of heavy vector-like quarks under $SU(2)_L \times U(1)_Y \times U(1)_{\mu-\tau} \times Z_2$.}
\label{tab:chargevq}
\end{table}
\end{center}

The relevant mass terms and the Yukawa interactions of the heavy vector-quarks are
\begin{eqnarray}\label{eq:Yukawa}
\mathcal{L}_{\text{VLQ}}&=&S \left(\overline{{\tilde D}}_R Y_{Qj} P_L d_j + \overline{{\tilde D}}_L Y_{Dj} P_R d_j + \overline{{\tilde U}}_R Y_{Qj} P_L u_j + \overline{{\tilde U}}_L Y_{Uj} P_R u_j\right)\\ \nonumber
&&+M_Q \overline{Q}_L \tilde Q_R + M_D \overline{{\tilde D}}_L D_R +  M_U \overline{{\tilde U}}_L U_R ~+~ \hc ~,
\end{eqnarray}
After integrating out the heavy vector-like quarks, Eq.~(\ref{eq:Yukawa}) induces an effective coupling of $Z^\prime \bar{d}_i d_j$ with the following form
\begin{eqnarray}
g^\prime\left(\bm{L}^d_{ij} \bar{d}_i \gamma^\mu P_L d_j Z^\prime_\mu + \bm{R}^d_{ij} \bar{d}_i \gamma^\mu P_R d_j Z^\prime_\mu\right)~,
\label{eq:Yukawa-eff}
\end{eqnarray}
where the Yukawa-like matrices $\bm{L}^d_{ij}$ and $\bm{R}^d_{ij}$ depend on $m_{Q,D}$ and $Y_{Q,D}$ as~\cite{Altmannshofer:2014cfa}
\begin{align}
\bm{L}^d_{ij}= \frac{v^2_S}{2 M_Q^2}\left(Y_{Qi} Y^*_{Qj}\right)~,\quad
\bm{R}^d_{ij} = -\frac{v^2_S}{2 M_D^2}\left(Y_{Di} Y^*_{Dj}\right)~.
\end{align}
Obviously, they are hermitian matrices. Without loss of generality, all of components can be taken as real. We further simplify these matrices by only keeping the flavor-diagonal components and the components which are related to $bs$ transition. Thus, $\bm{L}^d_{ij}$ and $\bm{R}^d_{ij}$ take the form
\begin{align}
\bm{L}^d_{ij} = \frac{v^2_S Y^2_{Q}}{2 M^2_{Q}}
\begin{pmatrix}
1 & 0 & 0\\
 0& 1 & 1 \\
 0& 1 & 1
\end{pmatrix}~,\quad
\bm{R}^d_{ij} = -\frac{v^2_S Y^2_{D}}{2 M^2_{D}}
\begin{pmatrix}
1 & 0 & 0\\
 0& 1 & 1 \\
 0& 1 & 1
\end{pmatrix}~,
\label{eq:texture}
\end{align}

Considering the best-fit values of Wilson coefficients for $R_{K^{(*)}}$ anomaly in Ref.~\cite{Aebischer:2019mlg}, we found that current data implies $C_9^{\prime\,\mu}\approx 0$. For simplicity, we neglect $C_9^{\prime\,\mu}$ contribution in the later discussions by setting $m_D$ decoupled from the spectrum. After above manipulations, from Eqs.~(\ref{eq:Yukawa-eff}) and~(\ref{eq:texture}), one obtains following effective Hamiltonian
for $b\to s\mu\mu$ decays
\begin{eqnarray}
\mathcal{H}^{bs\mu\mu}_{\rm eff} = \frac{g^{\prime 2} \bm{L}^d_{23}}{M^2_{Z^\prime}} \bar{s} \gamma^\mu P_L b \bar{\mu}\gamma_\mu \mu~,
\end{eqnarray}
for heavy $Z^\prime$, and corresponding Wilson coefficient $C_9^\mu$ with muons reads
\begin{align}
C_9^\mu = \frac{g^{\prime~2}\bm{L}^d_{23}}{M^2_{Z^\prime}}\frac{\sqrt{2}\pi} {G_F V_{tb}V_{ts}^*\alpha_{em}}
= \frac{\pi }{\sqrt{2}G_F V_{tb}V_{ts}^*\alpha_{em}}\left(\frac{Y_{Q}}{M_Q}\right)^2~.
\label{eq:wilson1}
\end{align}
Taking typical values $G_F = 1.166\times10^{-5}~\GeV^{-2}$,  $V_{tb}V_{ts}^*\approx -4.058\times10^{-3}$, Eq.~(\ref{eq:wilson1}) yields
\begin{align}
C_9^\mu \approx -6.43\times10^9 \left(\frac{Y_{Q}}{M_Q}\right)^2 = -1.61\left(\frac{Y_{Q}^2 }{0.1}\right)\left(\frac{20 \TeV}{M_Q}\right)^2~.
\end{align}
To interpret the $R_{K^{(\ast)}}$ anomaly, the $1\sigma$ range $C_9^\mu\in[-1.10,-0.79]$ is required \cite{Aebischer:2019mlg}. It is clear that the coefficient $C_9^\mu$ only depends on parameters $Y_Q$ and $M_Q$, leaving the $U(1)_{L_\mu-L_\tau}$ parameters $m_{Z'}$ and $g'$ free to choose. In the following discussion, we fix $Y_Q=0.122$ and $M_Q=10~\TeV$ to acquire the best fit value of $C_9^\mu\approx-0.95$ \cite{Aebischer:2019mlg}.

\subsection{Constraints}\label{sec:CT}

Although interpretation of $R_{K^{(*)}}$ in above section do not depends on $M_{Z'}$ and $g'$,
the relevant parameter space of $Z'$ is constrained by the following experiments:
\begin{itemize}
  \item \textbf{Moun $g-2$ and neutrino trident production}

In our model, the $Z^\prime$ contribution to the muon magnetic moment anomaly is given as
\begin{align}
\Delta{a_\mu}\equiv \frac{(g-2)_\mu}{2}=\frac{g^{\prime2}}{8\pi^2}\int^1_0 \frac{2x^2 (1-x)}{x^2+(M_{Z^\prime}/M_\mu)^2(1-x)}dx~,
\end{align}
However, the allowed parameter space is tightly constrained by neutrino trident production \cite{Altmannshofer:2014pba}, i.e., $\nu_\mu N\to \nu_\mu N\mu^+\mu^-$ process. In the heavy $Z^\prime$ case, the normalized cross section expressed as~\cite{Altmannshofer:2014cfa}
\begin{align}
\frac{\sigma}{\sigma_{\rm SM}}
\simeq \frac{1+\big(1+4s_W^2 + 8 \frac{g^{\prime2}}{g^2_2}\frac{M^2_W}{M^2_{Z^{\prime}}}\big)^2}{ 1+(1+4s_W^2 )^2}~.
\end{align}
In this paper, we consider the CCFR measurement $\sigma/\sigma_{\rm SM}=0.82\pm0.28$~\cite{Mishra:1991bv}.

\item \textbf{$B_s-\bar{B}_s$ mixing}

In our model, the flavor changing vortex $Z^\prime s b$ leads to tree level $B_s$ mixing via $Z^\prime$ exchange. In addition, Yukawa interactions in Eq.~(\ref{eq:Yukawa}) also contribute to $B_s$ mixing via box diagram, where singlet scalar and heavy vector-like quark running in the loop. The modification of mixing amplitude for heavy $Z^\prime$ yield~\cite{Altmannshofer:2014cfa}
\begin{eqnarray}
\frac{\Delta{M_{12}}}{M_{12}^\text{SM}} &\simeq& \left(\frac{Y_Q}{M_Q}\right)^4 \left[\left(\frac{M_{Z^\prime}}{g^\prime}\right)^2+\frac{M^2_Q}{16\pi^2}\right]
\times\left[\frac{g^4_2}{16\pi^2}\frac{1}{M^2_W}(V_{tb}V^*_{ts})^2S_0\right]^{-1}~,
\label{eq:bsmix1}
\end{eqnarray}
with $S_0\approx2.3$. Current experimental measurement allows $\Delta M_{12}/ M_{12}^{\text{SM}}\lesssim15\%$~\cite{Charles:2015gya}.

\item \textbf{Branching ratio for $t\to cZ^\prime$}

This is also induced by the left-handed $t\to c$ current, which is related to the left-handed $b\to s$ current by SU(2)$_L$ symmetry. Provided $Y_{Qt}\sim Y_{Qb}$, $Y_{Qc}\sim Y_{Qs}$~\cite{Fuyuto:2015gmk},
the branching ratio is
\begin{align}
\BR(t\to cZ^\prime)_{\rm LH}
\simeq \frac{(1-x_{Z^\prime})^2(1+2x_{Z^\prime})v^2}{8(1-x_W)^2(1+2x_W)} \left( \frac{Y_Q}{M_Q}\right)^4 \left(\frac{M_{Z^\prime}}{g^\prime}\right)^2, \label{eq:t2cZp}
\end{align}
where $x_W\equiv M_W^2/M_t^2$, $x_{Z^\prime} \equiv M_{Z^\prime}^2/M_t^2$
and we have set $M_c^2/M_t^2$, $M_b^2/M_t^2 \to 0$.
The decay $t\to cZ^\prime$ followed by $Z^\prime\to\ell^+\ell^-$ ($\ell=\mu,\tau$) can be searched for in $t\bar t$ events at the LHC.
It is similar to $t\to qZ$ ($q=u,c$) decay, which has been searched for
by the ATLAS~\cite{Aad:2015uza} and CMS~\cite{Chatrchyan:2013nwa} experiments
using $t\bar t \to Zq+Wb$ with leptonically decaying $Z$ and $W$, resulting
in a final state with three charged leptons. Reinterpreting the CMS limits for $t\to cZ$
to the case for $t\to cZ^\prime$ by a simple scaling of $Z$ and $Z^\prime$ decay branching ratios into the charged leptons ($\ell=e,\mu$), Ref.~\cite{Fuyuto:2015gmk} found  $\BR (t\to cZ')\lesssim10^{-4}$.

\item \textbf{Branching ratio for $Z\to4\ell$}

In our model, the $Z\to 4\ell$ decay will receive a significant contribution from $Z\to \mu^+\mu^-Z^\prime $ followed by $Z^\prime\to \mu^+\mu^-$ for the case of $M_{Z^\prime}< M_{Z}$. The ATLAS~\cite{TheATLAScollaboration:2013nha} and CMS~\cite{CMS:2012bw,Sirunyan:2018nnz} collaborations both have set upper limits on the branching fraction of the $Z$ boson decay to four charged leptons. In particular ATLAS has set an upper limit on
$\BR(Z\to 4\mu)=(4.2\pm0.4)\times10^{-6}$ with the combined 7
TeV and 8 TeV dataset~\cite{TheATLAScollaboration:2013nha}.
Using $77.6~\fb^{-1}$ data  at 13 TeV, CMS recently sets a more stringent upper limits of $10^{-8}\sim10^{-7}$ on the branching ratio BR$(Z\to Z'\mu\mu)$BR$(Z'\to\mu\mu)$ \cite{Sirunyan:2018nnz}. In this work, we adopt the dedicated limits on $L_\mu-L_\tau$ model provided by CMS \cite{Sirunyan:2018nnz}.

\item \textbf{LHC $Z^\prime$ constraints on dilepton final state.}

In our model, $Z^\prime$ boson will be produced at LHC through the flavor
conserving process $q\bar{q}\to Z^\prime$ and flavor violating process $b\bar{s}\to Z^\prime$ (and its conjugate process). Therefore, searches of heavy resonance in the dimuon final state by ATLAS~\cite{ATLAS:2016cyf,Aaboud:2017buh} and CMS~\cite{CMS:2016abv} tightly constrain the parameter space. In particular, ATLAS~\cite{ATLAS:2016cyf,Aaboud:2017buh} has set a $95\%$ C.L. upper limit on $\sigma(pp\to Z^\prime+ X) \BR(Z^\prime\to\mu^+\mu^-)$ in
the 150 GeV $\lesssim M_{Z^\prime} \lesssim$ 5 TeV mass range, with the 13
TeV and $\sim$ 13 fb$^{-1}$ dataset.
\end{itemize}
Other experiments, such as $\tau$ decays, are less strict than the above ones \cite{Altmannshofer:2014cfa}, we thus do not take into account in this paper. The above mentioned constraints are shown in figure~\ref{Fig:gmz1}.

\section{Dark Matter Phenomenology}\label{sec:DM}

In this section, we further investigate the phenomenology of Majorana fermion DM $N$ for high mass $Z'$. The model is implemented in {\tt FeynRules} \cite{Alloul:2013bka}. The calculation of DM relic density and DM-nucleon scattering cross section are performed with the help of {\tt micrOMEGAs} \cite{Belanger:2014vza}. The possible annihilation channels in this gauged $U(1)_{L_\mu-L_\tau}$ scotogenic model are listed in the following.
\begin{itemize}
  \item $NN\to \ell^+\ell^-, \nu_\ell\nu_\ell(\ell=e,\mu,\tau)$ is mediated by the inert scalar doublet $\eta$ via the Yukawa coupling $f_\ell$. Tightly constrained by LFV, $f_\ell\lesssim0.01$ is usually needed for electroweak scale $N$ and $\eta$ \cite{Kubo:2006yx}, thus contributions of this channel are negligible.
  \item $NN\to Z^{'*}\to \ell^+\ell^-, \nu_\ell\nu_\ell (\ell=\mu,\tau)$ via the gauge coupling $g'$ provides a new $s$-channel process for DM annihilation. Different from the gauged $U(1)_{B-L}$ case \cite{BL}, this channel exclusively generates muon, tau leptons and neutrinos.
  \item $NN\to h^*/H_0^*\to W^+W^-,b\bar{b},\ldots$ via the Yukawa coupling $h_{e\mu}$, $h_{e\tau}$ is also a $s$-channel process. Previous study neglected this channel by assuming tiny mixing angle $\alpha$ \cite{Baek:2015fea}. In this paper, a not too small mixing angle $\alpha$ is considered.
  \item $NN\to Z' Z', Z'H_0, H_0H_0, hH_0$ are also possible if kinematically allowed. As shown latter, the $NN\to Z'Z', Z'H_0(\to Z'Z')$ channel is possible to interpret the AMS-02 positron excess.
\end{itemize}

\begin{figure}
\begin{center}
\includegraphics[width=0.45\linewidth]{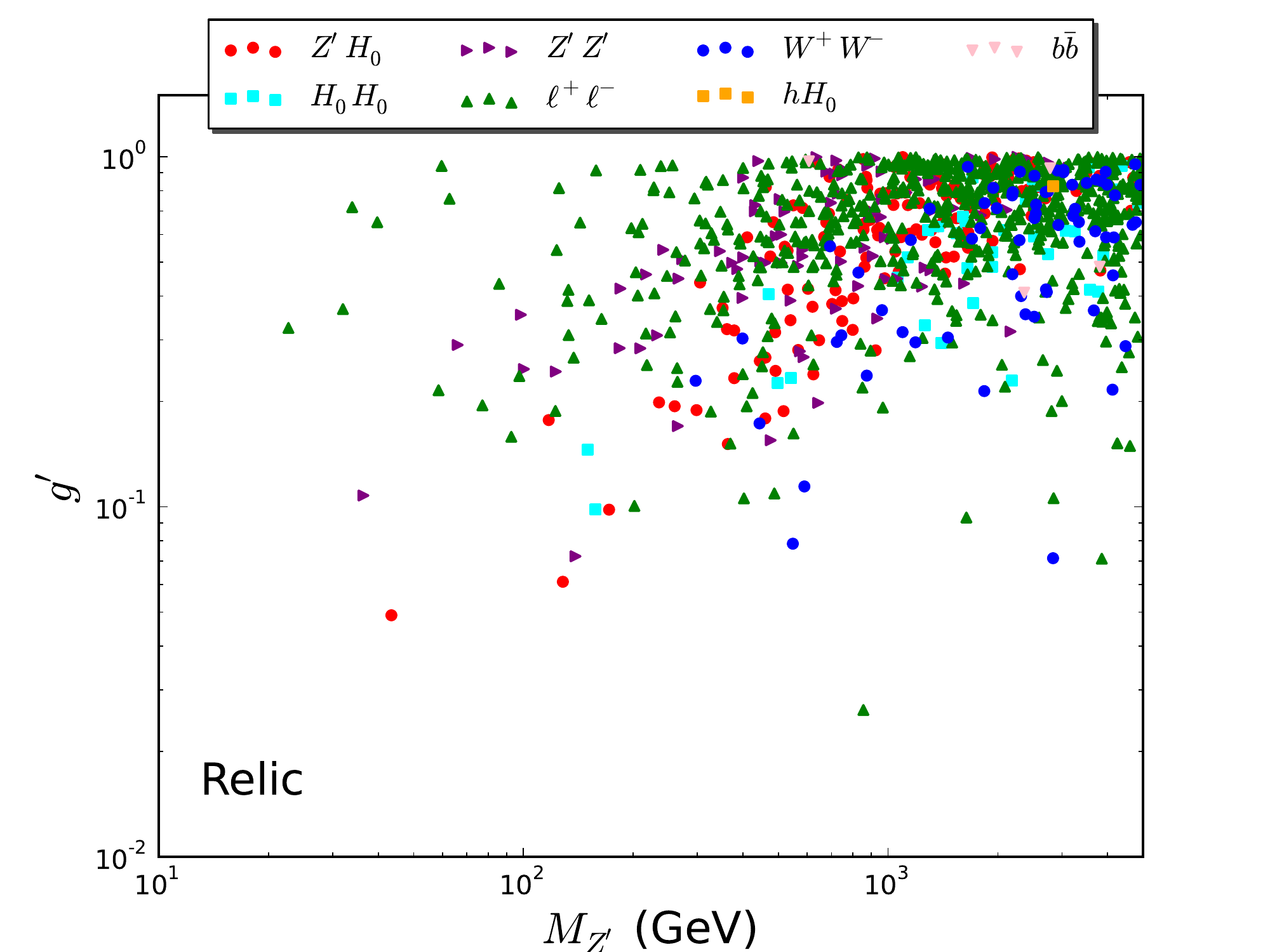}
\includegraphics[width=0.45\linewidth]{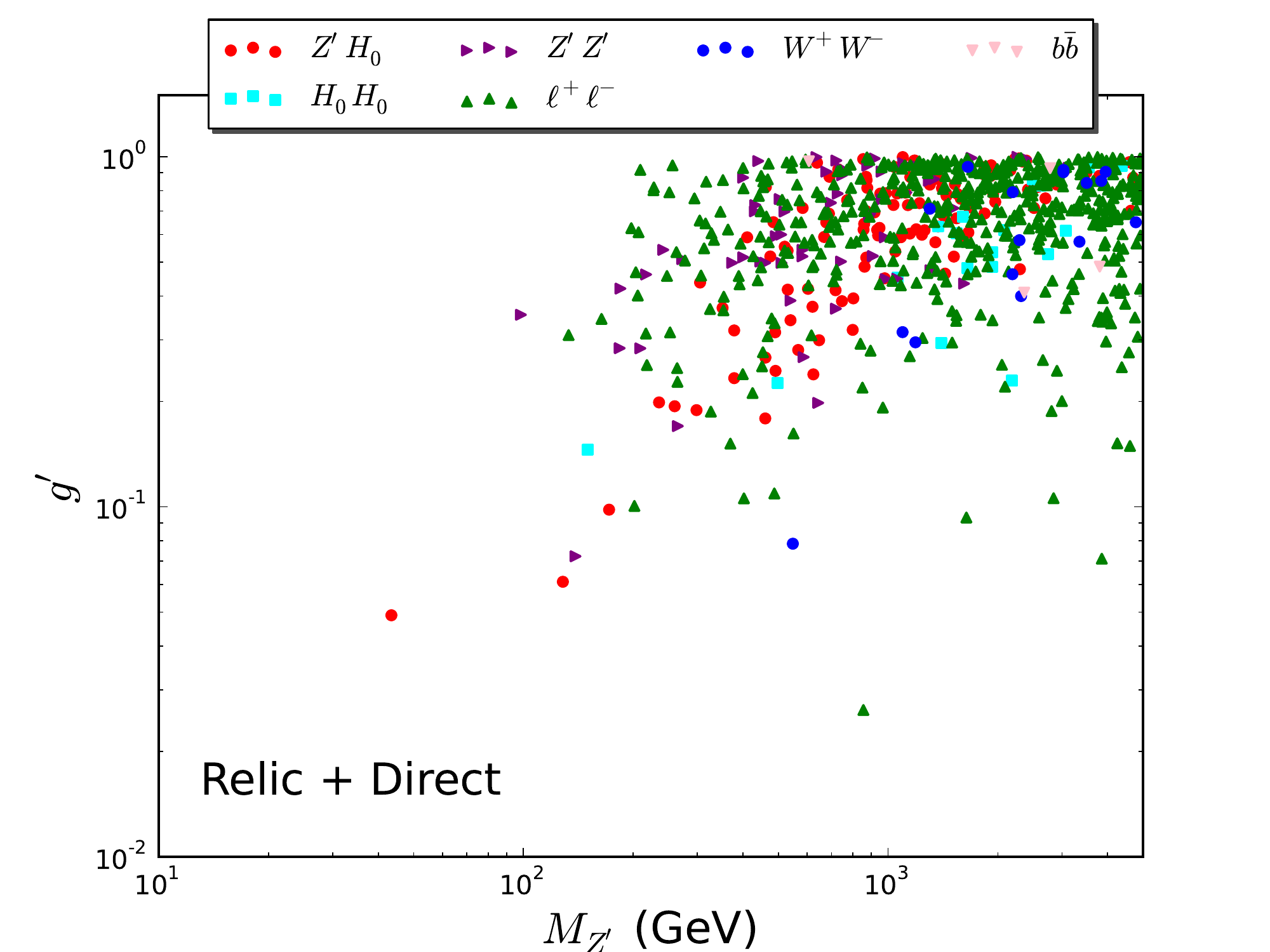}
\end{center}
\caption{Distribution of dominant annihilation channels in the $g'$-$M_Z'$ plane. Survived samples in left panel satisfy constraints from relic density, while those in right panel satisfy constraints from both relic density and direct detection.
\label{Fig:PR1}}
\end{figure}

To illustrate the effects of above various annihilation processes, we implement a random scan over the following parameter space
\begin{eqnarray}
g'\in[0.001,1]&,&M_{Z'}\in[10,5000]~\GeV, \\ \nonumber
\alpha\in[0.01,0.1]&,& M_{H_0} \in [0,\sqrt{4\pi} M_{Z'}/g'], \\ \nonumber
h_{e\mu,e\tau}\in[0,4\pi]&,& M_{ee,\mu\tau}\in[10,5000]~\GeV,
\label{eq:para}
\end{eqnarray}
and assign the dominant annihilation channel to the survived samples under constraints from relic density and direct detection. For relic density, we use the combined Planck+WP+highL+BAO $2\sigma$ value, i.e., $0.1153<\Omega h^2<0.1221$~\cite{Ade:2015xua}. As for direct detection, we adopt the combined limits provided by XENON1T~\cite{Aprile:2017iyp} and PandaX-II \cite{Cui:2017nnn}.
Meanwhile, to satisfy the observed neutrino oscillation parameters, we further require $|R|$ defined in Eq.~(\ref{eq:R}) in the interval $[0.4,0.5]$ with $\theta_R=\pi$ for simplicity \cite{Baek:2015fea}.

\begin{figure}
\begin{center}
\includegraphics[width=0.45\linewidth]{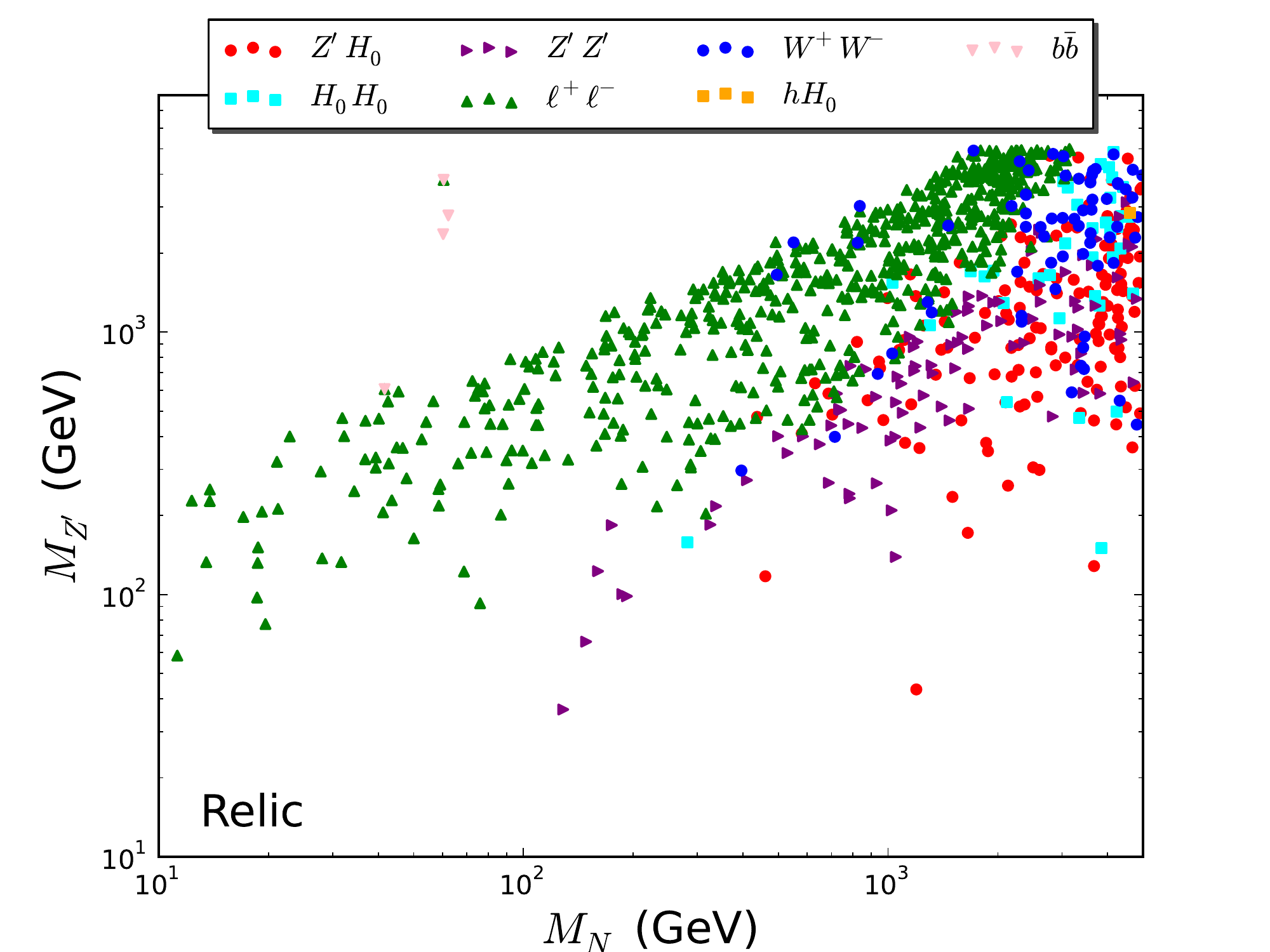}
\includegraphics[width=0.45\linewidth]{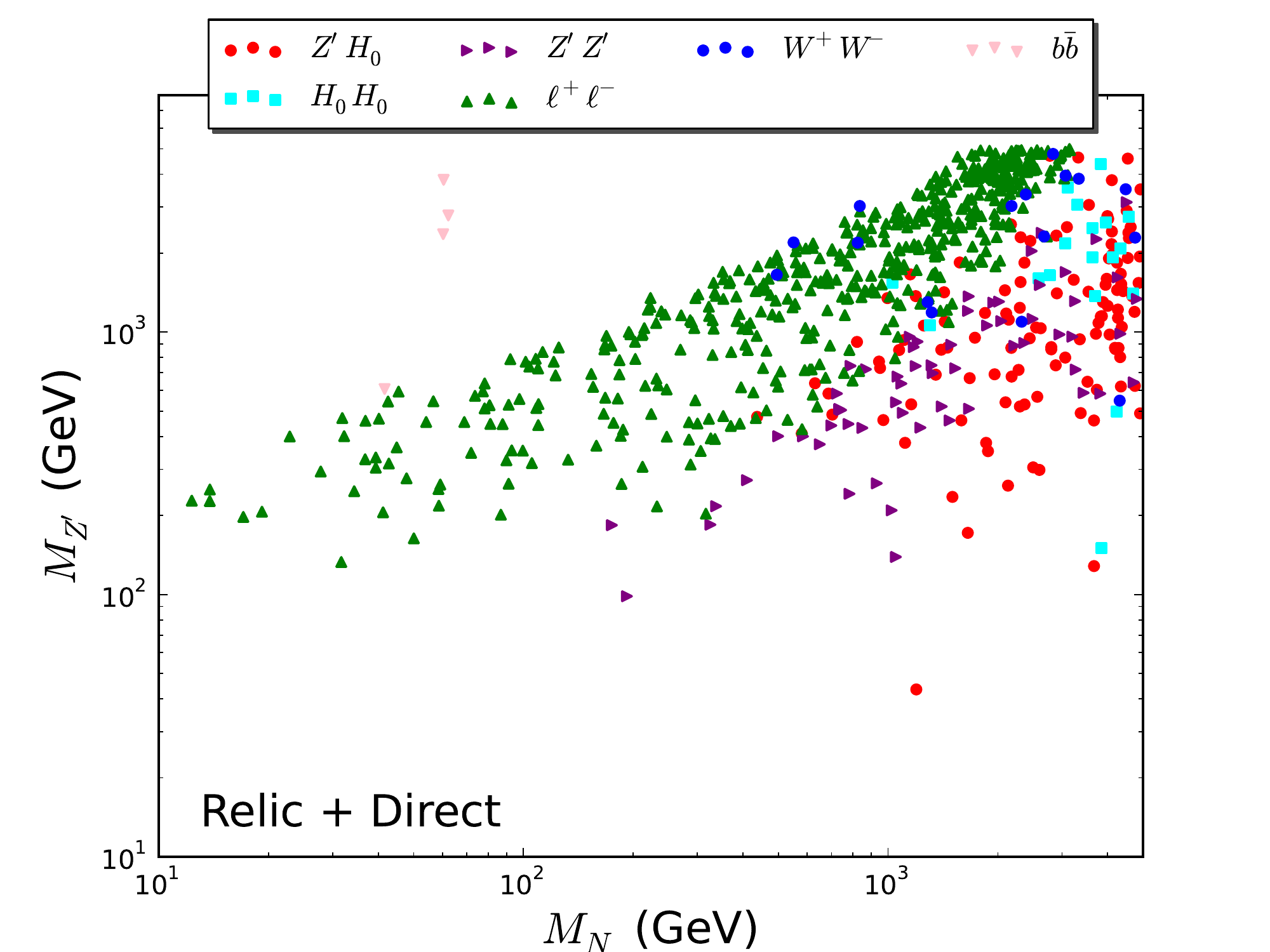}
\end{center}
\caption{Same as figure.~\ref{Fig:PR1}, but in the $M_{Z'}$-$M_N$ plane.
\label{Fig:PR2}}
\end{figure}

Due to the Majorana nature of $N$ DM, the DM-nucleon scattering cross section mediated by $Z'$ is suppressed, and is actually dominant by Higgs exchange. In this way, the spin-independent cross section is given by \cite{Okada:2010wd}
\begin{equation}
\sigma^\text{SI}=\frac{h_N^2 \mu_N^2 M_n^2 f_n^2}{2 \pi v^2} \sin^2 2\alpha
\left(\frac{1}{M_h^2}-\frac{1}{M_{H_0}^2}\right)^2,
\end{equation}
where $h_N=(V_{eN}V_{\mu N}+V_{\mu N}V_{e N}) h_{e\mu}+ (V_{eN}V_{\tau N}+V_{\tau N}V_{e N}) h_{e\tau}$ is the effective DM-$S$ coupling, $M_n\approx0.939~\GeV$ is the averaged nucleon mass, $\mu_N=M_n M_N/(M_n+M_N)$ is the DM-nucleon reduced mass, $f_n\approx 0.345$ is the nucleon matrix element. Clearly, the cross section is proportional to $\sin^22\alpha$, therefore a smaller mixing angle $\alpha$ is also preferred by direct detection.

In figure.~\ref{Fig:PR1} and \ref{Fig:PR2}, we depict the distribution of survived samples in the $g'$-$M_{Z'}$ and $M_{Z'}$-$M_N$ plane respectively. From figure~\ref{Fig:PR1}, we aware that correct relic density could be realized with $g'\gtrsim0.02$ and $M_{Z'}\gtrsim20~\GeV$ during our scan, but the direct detection would exclude those points with $M_{Z'}\lesssim100~\GeV$. Note that a little points dominant by $hH_0$ is survived under relic density, but such points are fully excluded by direct detection. Distributions of different annihilation channels in the $M_{Z'}$-$M_N$ plane as shown in figure~\ref{Fig:PR2} are clearer for different annihilation channels. The $NN\to Z'^*\to \ell^+ \ell^-$ channel is dominant in the region $M_{Z'}\sim 2 M_{N}$. When $M_{Z'}<M_N$ or $M_{H_0}<M_N$, the dominant channels become $NN\to Z'Z'$, $ZH_0$ and $H_0H_0$. For the $s$-channel Higgs portal dominant, the $NN\to b\bar{b}$ channel needs $M_N\sim M_h/2\approx60~\GeV$, while $NN\to W^+W^-$ channel requires $M_N\gtrsim500~\GeV$.

\begin{figure}
\begin{center}
\includegraphics[width=0.6\linewidth]{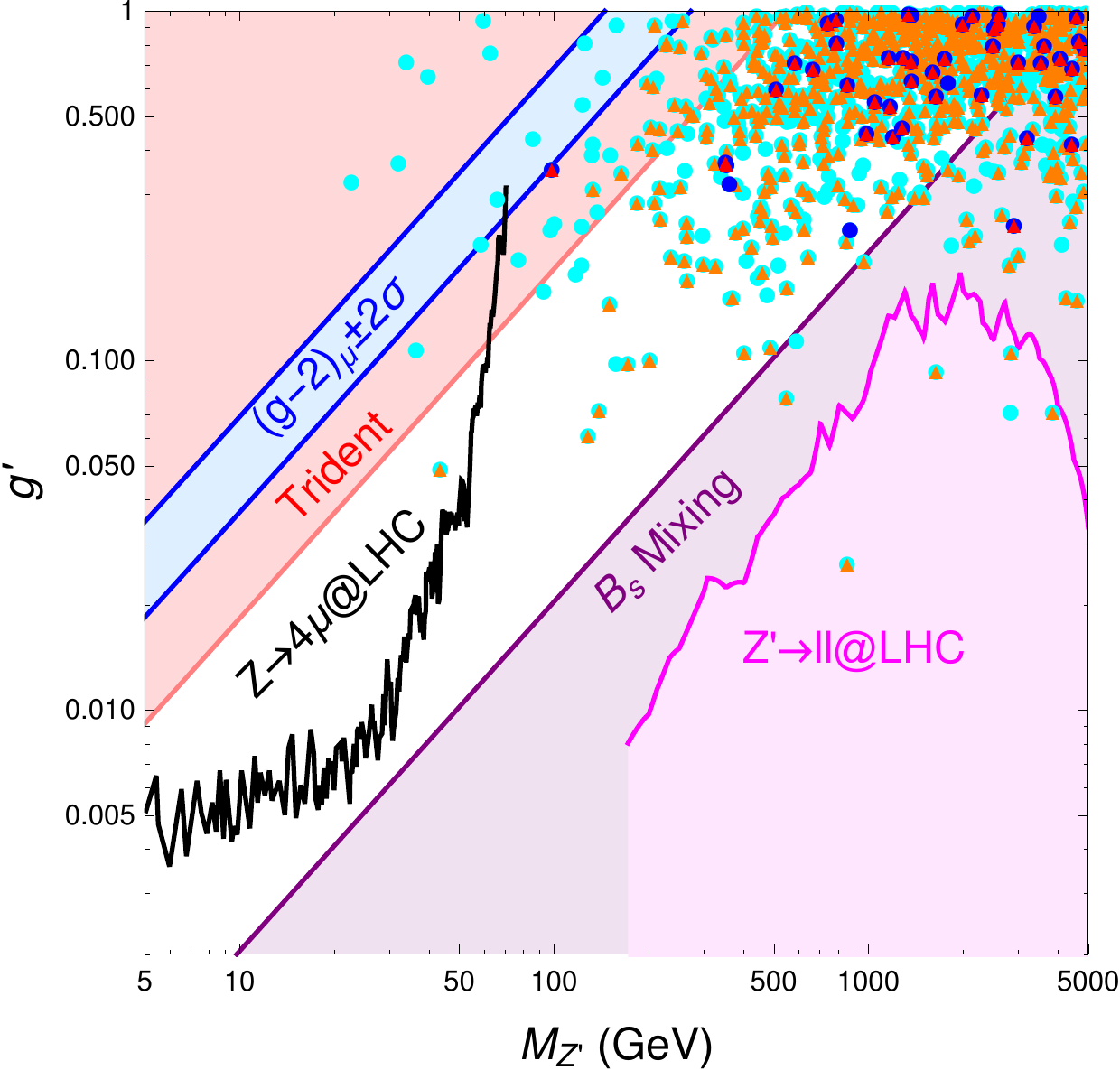}
\end{center}
\caption{Survived samples in the $g'$-$M_Z'$ plane with constraints from Sec.~\ref{sec:CT}. The constraint $t\to c Z'$ excludes $g'\lesssim4\times10^{-4}$, which is too small to show here. The cyan circle points satisfy relic density only, while the orange triangle points further satisfy  direct detection. Blue circle points satisfy relic density and neutrino oscillation, and the red triangle points further satisfy direct detection.
\label{Fig:gmz1}}
\end{figure}

In figure~\ref{Fig:gmz1}, we show combined results from relic abundance, direct detection, neutrino oscillation as well as various constraints on $Z'$ discussed in Sec.~\ref{sec:CT}. The neutrino trident production process has excluded the $(g-2)_\mu$ favor region, and set the most stringent  upper limit on $g'$ for $M_{Z'}\gtrsim50~\GeV$. For $M_{Z'}\lesssim50~\GeV$, the most stringent upper limits comes from $Z\to4\mu$ search at LHC. Meanwhile, the $B_s$ mixing has set an lower limit on $g'$.
It is clear that a few survived red triangle samples are not excluded by neutrino trident production and $B_s$ mixing. Hence, viable parameter space is obtained to explain DM, neutrino mass and $R_{K^{(*)}}$ anomaly simultaneously.

\section{AMS-02 Positron Excess}\label{sec:AMS}

\subsection{AMS-02 Positron Flux}\label{sec:po}

In this section, we discuss the AMS-02 positron excess and relevant constraints from indirect detections. Recently, the AMS Collaboration has released latest result of positron spectrum that extend the maximal measurement energy up to 1 TeV~\cite{Aguilar:2019owu}, which is used in our fitting. For a given model parameters in Eq.~(\ref{eq:para}), the positron flux can be expressed as
\begin{align}
 \Phi_{e^+}(f_{e^+},\phi_{e^+}^\odot, BF) = f_{e^+} \Phi_{e^+}^{\rm bkg}(\phi_{e^+}^\odot)+\Phi_{e^+}^{\rm DM}(\phi_{e^+}^\odot, BF),
\label{eq:frac}
\end{align}
where $f_{e^+}$ is the normalization factors which take into account the uncertainty of the astrophysical background and varying in the range $[0,~5]$ in our fitting. The fluxes of charged CR particles are periodically modulated according to the solar activity due to their interactions with the heliosphere magnetic field. The modulation is more important for low energy CR particles and can be described by using the force field approximation~\cite{Gleeson:1968zza}. In this approximation, the modulated spectrum $\Phi_{\rm mod}(E_k)$ and unmodulated one $\Phi(E_k)$ is related by following formula:
\begin{align}
\Phi_{\rm mod}(E_k)=\frac{(E_k+M)^2-m_{\rm CR}^2}{(E_k+m_{\rm CR}+|e|\phi^\odot)^2-m_{\rm CR}^2}
\Phi(E_k+|e|\phi^\odot),
\end{align}
where $m_{\rm CR}=m_e$ ($m_p$) for position (antiproton) flux, $\phi^\odot$ is the modulation potential, and $E_k$ is the observed kinetic energy.
Note that the force field approximation is an over simplified model.
The modulation potential $\phi^\odot$ is just an effective parameter that indicate the total effect in the solar modulation.
In fact, different CR particles would always require different modulation potentials.
We therefore consider the modulation potentials of positron and antiproton as two free parameters in the fitting.

We calculate the positron and antiproton flux resulted from DM annihilation using {\tt micrOMEGAs} \cite{Belanger:2014vza}. For the DM density distribution in the galactic halo, we have used the Navarro-Frenk-White (NFW) density profile~\cite{Navarro:1996gj} with the local density $\rho_\odot=0.4\mathrm{\,GeV\,cm}^{-3}$.
The background fluxes are obtained by solving the diffusion equation for cosmic-ray particles using a widely used galactic cosmic-ray propagation model.
To take into account the convection/reacceleration effect and the complex electron energy losses during the diffusion, we adopt the public code {\tt GALPROPv54}~\cite{Moskalenko:1997gh,Strong:1998pw}. The relevant parameters for these process, such as the diffusion coefficient and the convection velocity, ought to be determined by fitting to the B/C and proton data. Here we choose the these parameters following the diffusion + convection (DC) case in Ref.~\cite{Lin:2014vja} to derive both the secondary positron and antiproton fluxes. In all, we set the diffusion coefficient $D(R)=1.95\times10^{28}(R/4.71\mathrm{\,GV})^{0.51}\mathrm{\,cm^2\,s^{-1}}$, the gradient of covection velocity $\mathrm{d}V/\mathrm{d}z=4.2\mathrm{\,km\,s^{-1}\,kpc^{-1}}$
and the proton injection with a power-index 2.336.
The $\chi^2$ function is defined as
\begin{align}
\chi^2_{e^+}(f_{e^+},\Phi_{e^+}^\odot, BF)=\sum_i\frac{\left[\Phi_{e^+,i}(f_{e^+},\Phi_{e^+}^\odot, BF)-\Phi_{e^+,i}^{\rm AMS}\right]^2}{(\sigma_{e^+,i}^{\rm AMS})^2},
\label{eq:chipo}
\end{align}
where $i$ runs over all the data points. $\Phi_{e^+,i}$ and $\sigma_{e^+,i}^{\rm AMS}$ are respectively the relevant observables (positron fraction in this case) and corresponding experimental errors (stat+syst) taken from~\cite{Aguilar:2014mma,Aguilar:2016kjl}.

It has been known that in order to fit AMS-02 positron data, a large enhancement with $BF\sim\calO(10^3)$ is required for annihilation cross section in the Galaxy with $v\sim 10^{-3}$ than that in the freeze-out temperature with $v\sim 10^{-1}$. In addition, the annihilation final states should be leptophilic to avoid antiproton constraint~\cite{Lin:2014vja}. Even so, this scenario is still challenged by limits from extragalactic $\gamma$-ray background~\cite{Ackermann:2014usa} and CMB observations~\cite{Galli:2011rz,Finkbeiner:2011dx,Ade:2015xua,Slatyer:2015jla}. We will discuss these constraints in more detail in section~\ref{sec:constraints}. Here we first illustrate how to obtain a large $BF$ in a consistent way in our model. The two common methods widely used to simultaneously realize the correct relic abundance and a large $BF$ are so-called Breit-Wigner mechanism~\cite{Feldman:2008xs,Ibe:2008ye,Guo:2009aj} and Sommerfeld enhancement~\cite{Hisano:2003ec,Hisano:2006nn,ArkaniHamed:2008qn}. In the former case, two DM particles annihilate via the s-channel exchange of a heavy mediator, then the annihilation cross section are resonantly enhanced when mediator mass is close to twice of DM mass. Notably, it has been shown that Breit-Wigner mechanism can potentially relax the tension between positron excess and CMB observations due to the evolution of velocity dependent annihilation cross section at different cosmic epochs (freeze-out, recombination and present)~\cite{Xiang:2017jou}. Unfortunately, this mechanism has less effect on our model. Based on the discussion in Sec.~\ref{sec:DM}, the only important $s$-channel annihilation is $NN\to Z^{'*}\to \ell^+\ell^-$, which is $p$-wave suppression in the Galaxy since $N$ is Majorana DM. As a consequence, annihilation cross section is not large enough even in the resonance regions.

\begin{figure}
\begin{center}
\includegraphics[width=0.5\linewidth]{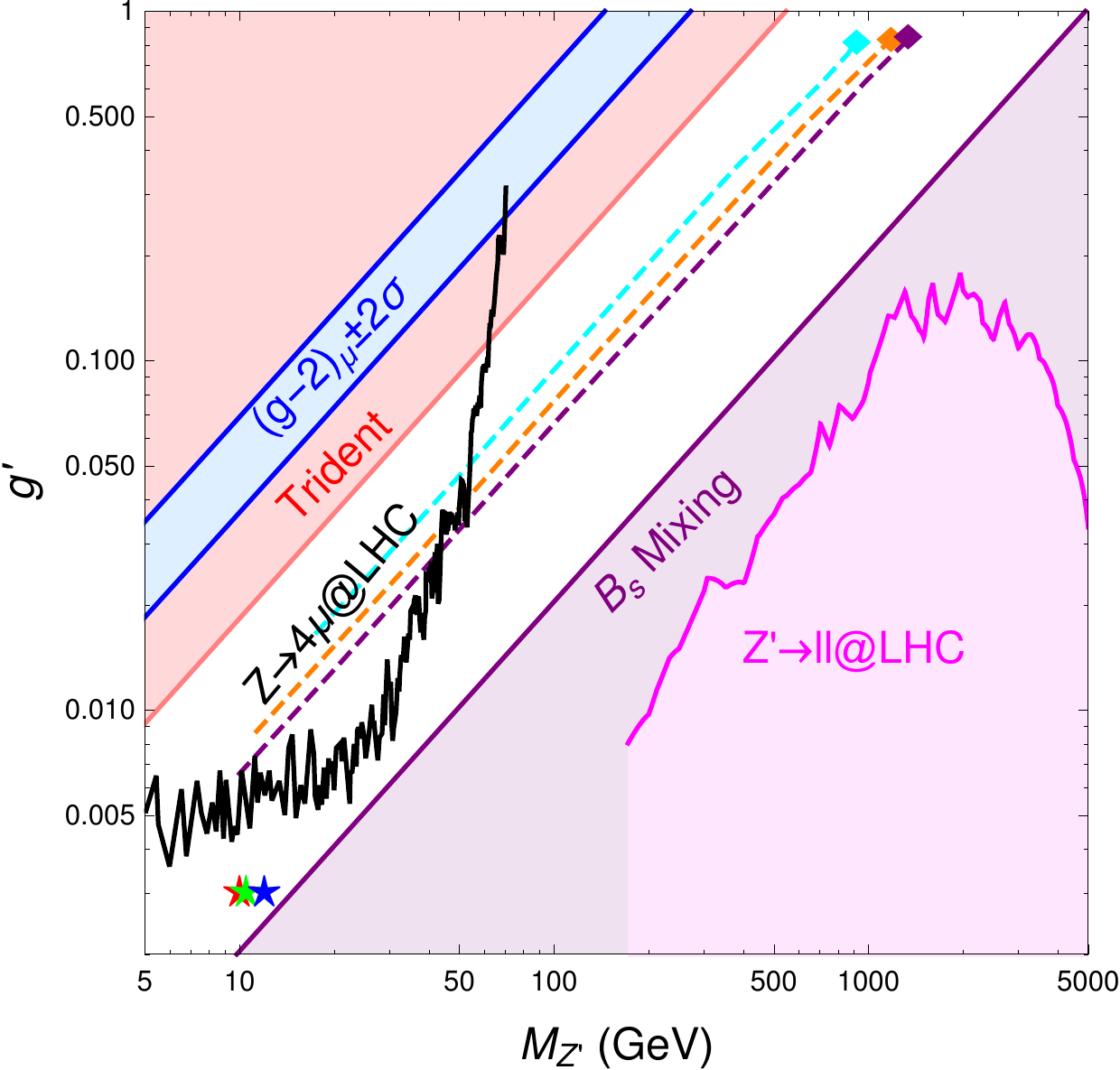}
\end{center}
\caption{The benchmarks in our model to fit AMS-02 positron data. Here cyan, orange and purple diamonds (red ,green and blue stars) respectively corresponding to $M_N =1,~1.5,~2$ TeV for $NN\to Z^\prime Z^\prime$ ($NN\to Z^\prime H_0$) annihilation channel. For completeness, various limits in figure~\ref{Fig:gmz1} are shown.}
\label{Fig:gmz2}
\end{figure}

We therefore focus on Sommerfeld enhancement in the $s$-wave DM annihilation. This mechanism is due to the loop correction of annihilation cross section with
exchange infinite number of vector or scalar mediators. In the non-relativistic limit, the velocity-dependent correction to DM annihilation can be computed by
numerically solving the radial Schr\"{o}dinger equation with the attractive spherically symmetric Yukawa potential $V(r)=-\alpha^\prime e^{-M^\prime}/r$, here $\alpha^\prime$
and $M^\prime$ respectively denote coupling and mass of mediator. The Sommerfeld enhancement factor $S_E$ then evaluated by the radial wave function at the origin,
$|\psi_k(0)|^2$. In following, we will use the semi-analytic formula introduced in Ref.~\cite{Cassel:2009wt,Slatyer:2009vg} for a illustration. Both $Z^\prime$ and $H_0$ can serve as mediators in our model, with corresponding annihilation channel are respectively $NN\to Z^\prime Z^\prime \to 2\ell^+2\ell^-+2\nu_\ell2\nu_\ell$ and $NN\to Z^\prime H_0(\to Z'Z') \to 3\ell^+3\ell^-+3\nu_\ell3\nu_\ell$ with $\ell=\mu,\tau$. However, annihilation channel $NN\to Z^\prime Z^\prime$ is difficult to give desired $BF$ through Sommerfeld enhancement and leaving $NN\to Z' H_0$ as only available channel, which is due to the fact that for $NN\to Z' Z'$ channel, its Sommerfeld enhancement factor determined by relic abundance and by positron excess are incompatible.

\begin{figure}
\begin{center}
\includegraphics[width=0.45\linewidth]{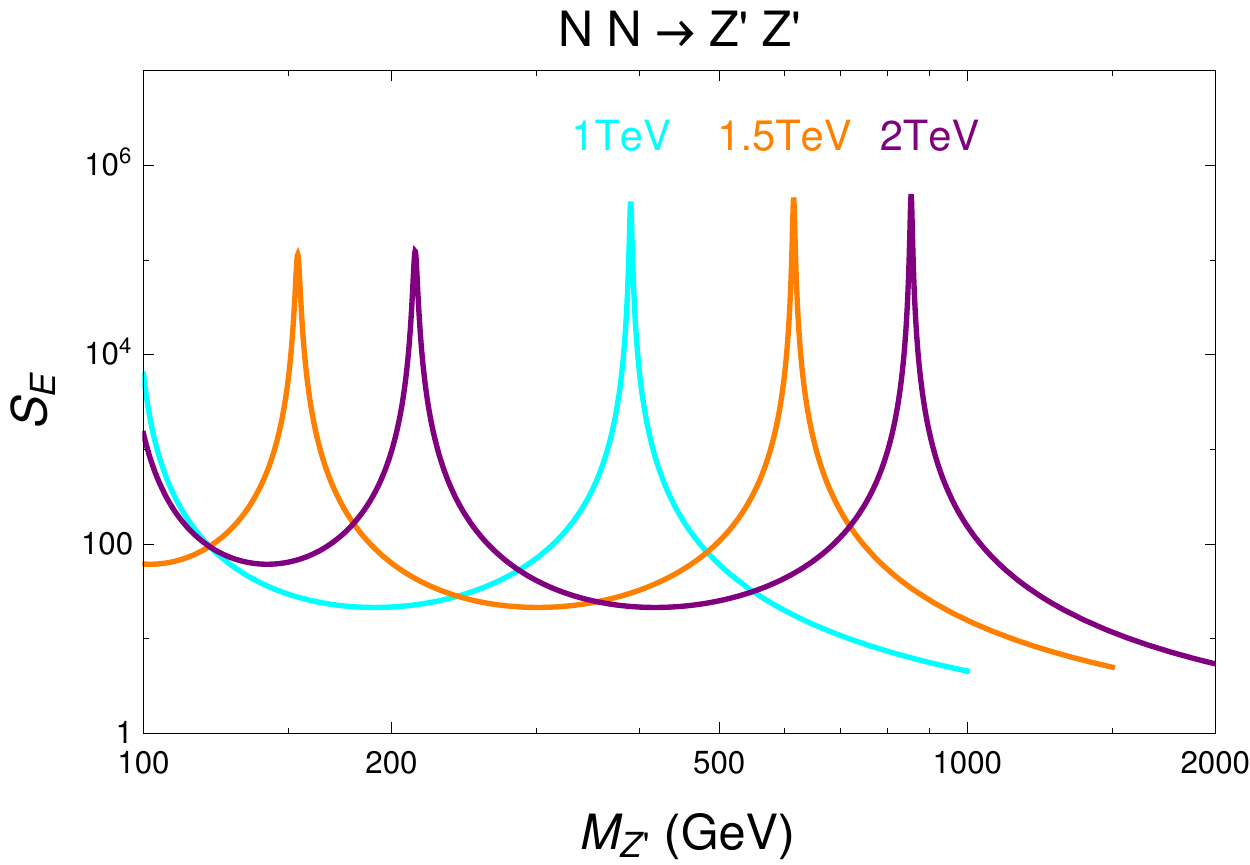}
\includegraphics[width=0.45\linewidth]{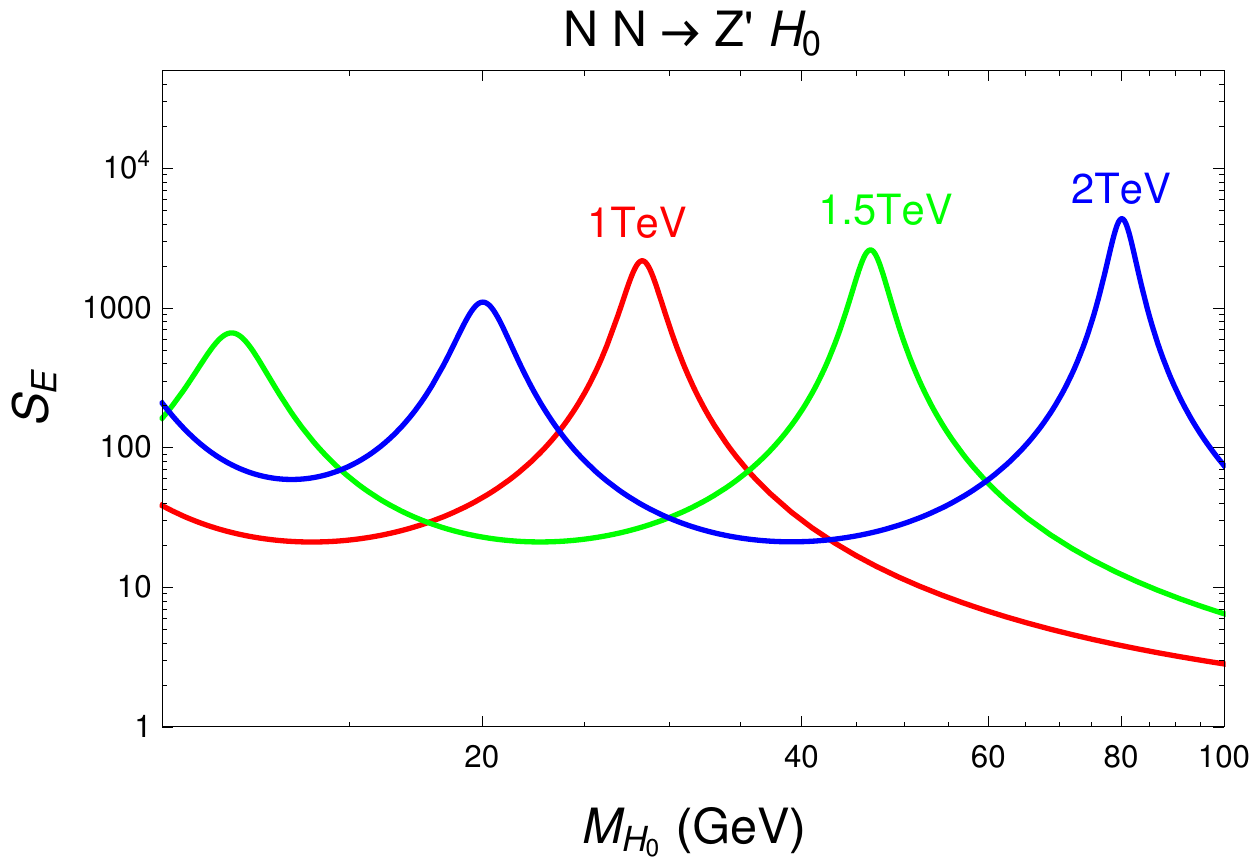}
\end{center}
\caption{The Sommerfeld enhancement factor $S_E$ in our model as a function of mediator mass $M_{Z^\prime}$ (left) and $M_{H_0}$ (right) for the benchmark parameters in table~\ref{tab:benchzz} and table~\ref{tab:benchzh}.}
\label{Fig:SE}
\end{figure}

To see this, we plot contours of correct relic abundance on the $g'-M_{Z'}$ plane for benchmark masses $M_N=1,~1.5,~2$ TeV in figure~\ref{Fig:gmz2} with combine various constraints from figure~\ref{Fig:gmz1}. Since annihilation cross section of $NN\to Z' Z'$ channel scales as $g'^4M^2_N/M^4_{Z'}$, relic abundance is entirely fixed by ratio $g'/M_{Z'}$ for a given $M_N$. Thus three curves presented just as straight line in the figure, and $Z'$ masses below $50$ GeV are excluded by $Z\to4\mu$ LHC direct search. We then select largest $(g',~M_{Z'})$ point in each curve as benchmark (cyan, orange and purple diamonds), their values are listed in table~\ref{tab:benchzz}. Corresponding $S_E$ curves for three benchmark $g'$ are shown in left panel of figure~\ref{Fig:SE}. Comparing the fitting values of $\langle\sigma v\rangle_{BF}\equiv BF\times\langle\sigma v\rangle_0$ in table~\ref{tab:benchzz} with $S_E$  values in the figure for the same $M_{Z'}$, it obviously fails to satisfy the requirement $S_E\simeq BF$. We further evaluate all of $(g',~M_{Z'})$ in relic abundance curves, none of them can match above condition. It boils down to the fact that both relic abundance and Sommerfeld enhancement factor $S_E$ share the same coupling $g'$ for given $M_N$ and $M_{Z'}$, which is hardly to tune its value to satisfy two requirements simultaneously.
 
As a consequence, we appeal to another annihilation channel $NN\to Z^\prime H_0$. This channel has advantage that its relic abundance and $S_E$ depend on parameters $(g',~h_N,~M_{Z'},~M_{H_0})$, which does not suffer from tight correlation between relic abundance and $S_E$ with such more degree of freedom. Similarly, we presented three benchmarks of $NN\to Z^\prime H_0$ channel in figure~\ref{Fig:gmz1} (red ,green and blue stars) and in table~\ref{tab:benchzh} for the same $M_N$. Here we take $g^\prime=3\times10^{-3}$ which is compatible with the LHC $Z\to4\mu$ direct search limit for light $Z^\prime$, and $M_{H_0}$ for each benchmark has been chosen such that $S_E\simeq BF$, as is shown in the right panel of figure~\ref{Fig:SE}.

\begin{figure}
\begin{center}
\includegraphics[width=0.45\linewidth]{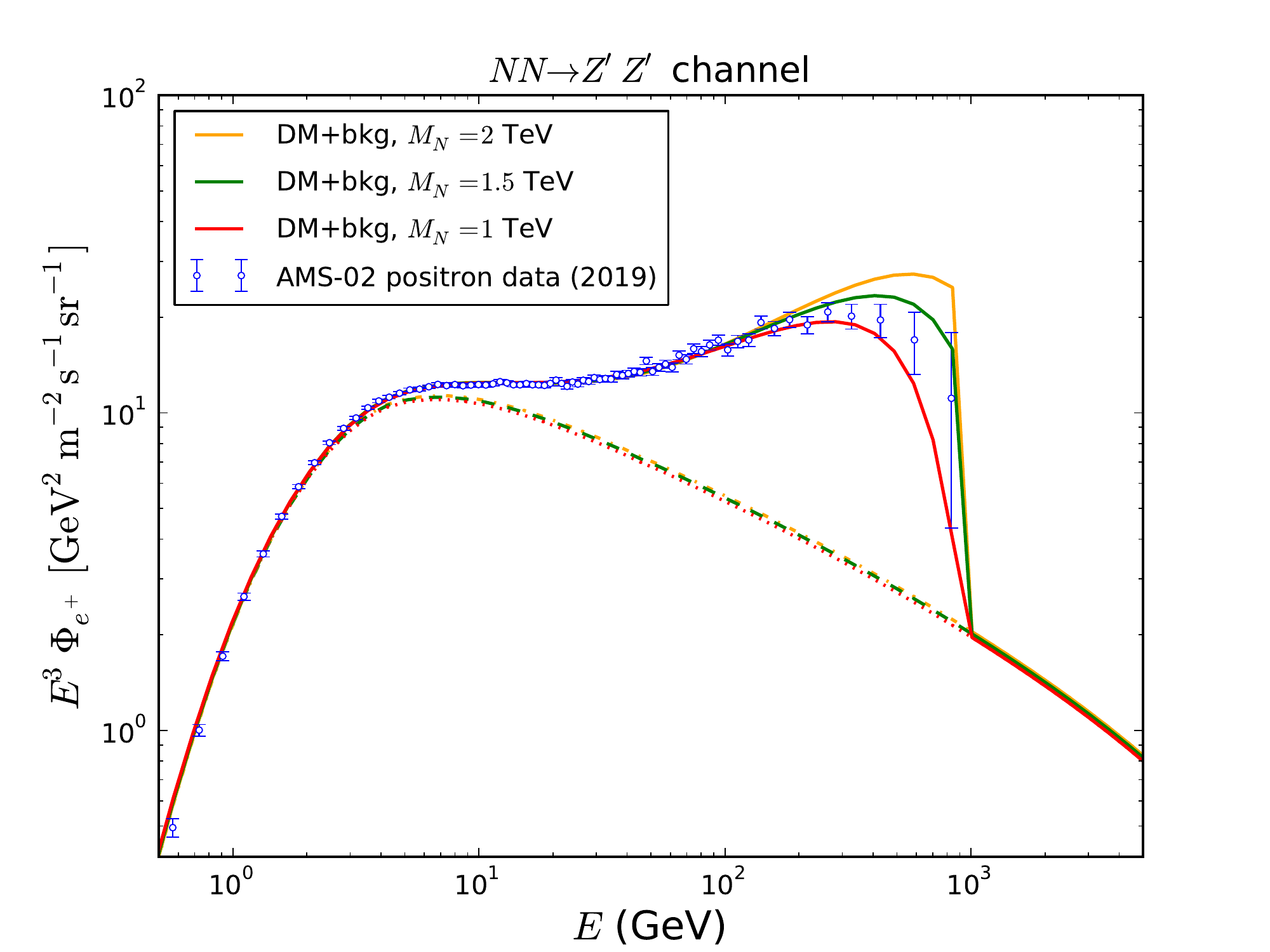}
\includegraphics[width=0.45\linewidth]{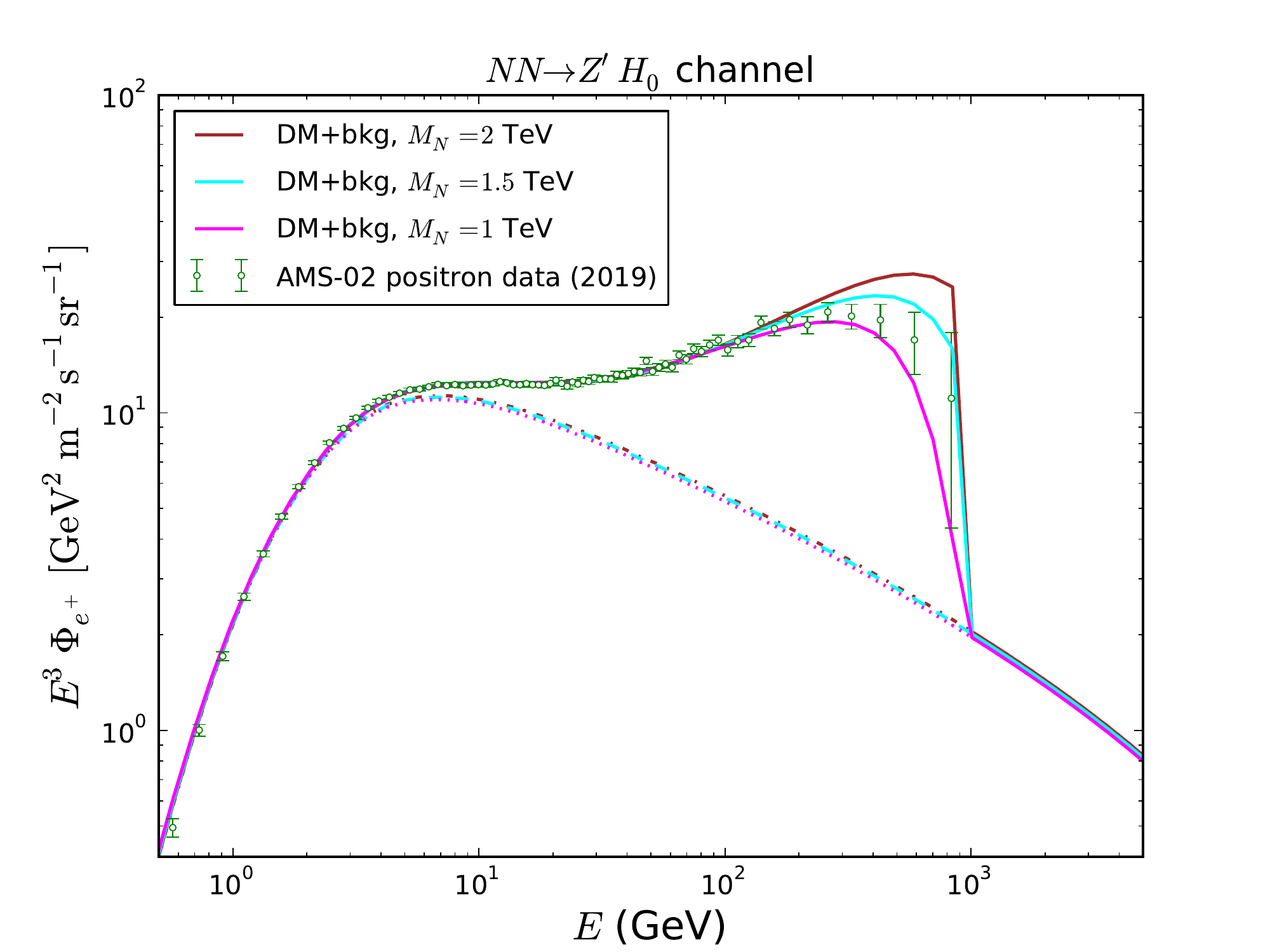}
\end{center}
\caption{The positron fluxes predicted by benchmarks of $NN\to Z'Z'$ (table~\ref{tab:benchzz}) and $NN\to Z'H_0$ (table~\ref{tab:benchzh}) annihilation channels with the fitting results of AMS-02 data.}
\label{Fig:po}
\end{figure}

The positron flux predicted by benchmarks in table~\ref{tab:benchzz} and~\ref{tab:benchzh} are shown in figure~\ref{Fig:po} with AMS-02 data. We found that both $NN\to Z' Z'$ and $NN\to Z^\prime H_0$ annihilation channels can provide good fitting for $M_N$ in the range of $1-1.5$ TeV, while 2 TeV benchmark results in too much excess.  Despite only $NN\to Z^\prime H_0$ channel gives consistent interpretation in Sommerfeld enhancement scenario, from the phenomenological viewpoint, we still treat $NN\to Z' Z'$ channel as a valid candidate. The relevant constraints for both two channels are investigated in the next section, .

\begin{table*}
\begin{tabular}{|c|c|c|c|c|c|c|c|c|c|c|c|c|c|c|}
\hline
$M_N$ & $M_{Z^\prime}$ & $g'$ & $\Omega_{\rm DM} h^2$ & $\langle\sigma v\rangle_0$& $f_{e^+}$ & $f_{\bar{p}}$ & $\phi_{e^+}^\odot$ & $\phi_{\bar{p}}^\odot$ & $\langle\sigma v\rangle_{BF}$ & $\langle\sigma v\rangle_{\rm CMB}$ & $\chi^2_{\rm min}(e^+)$ & $\chi^2_{\rm min} (\bar{p})$ \\\hline
$1000$ & $918$  & $0.642$ & $0.1207$ & $6.49\times10^{-27}$ & $0.78$  & $1.28$ & $601$  & $1019$ & $1.12\times10^{-23}$ & $3.27\times10^{-24}$ & $90.56$ & $75.75$\\\hline
$1500$ & $1180$ & $0.675$ & $0.1197$ & $8.81\times10^{-27}$  & $0.80$ & $1.28$ & $612$  & $1019$ & $2.16\times10^{-23}$ & $4.91\times10^{-24}$ & $81.32$ & $75.75$ \\\hline
$2000$ & $1338$ & $0.703$ & $0.1177$ & $7.46\times10^{-27}$  & $0.81$ & $1.28$  & $620$ & $1019$ & $3.48\times10^{-23}$ & $6.57\times10^{-24}$ & $111.38$ & $75.75$ \\\hline
\end{tabular}
\caption{The DM information for benchmarks of $NN\to Z^\prime Z^\prime$ annihilation channel. Here we choose $h_N=0.5$ and $\langle\sigma v\rangle_0$ (in units of ${\rm cm}^3~{\rm s}^{-1}$) denotes the thermally averaged DM annihilation cross section at freeze out, $\langle\sigma v\rangle_{BF}\equiv BF\times\langle\sigma v\rangle_0$ denotes annihilation cross section in the Galactic halo which required by interpreting AMS-02 positron excess, and $\langle\sigma v\rangle_{\rm CMB}$ the annihilation cross section limited by CMB observation. All of masses in units of $\GeV$ and solar modulation in units of MV.}
\label{tab:benchzz}
\end{table*}

\begin{table*}
\begin{tabular}{|c|c|c|c|c|c|c|c|c|c|c|c|c|c|}
\hline
$M_N$ & $M_{Z^\prime}$ & $M_{H_0}$ & $h_N$ & $\Omega_{\rm DM} h^2$ & $\langle\sigma v\rangle_0$ & $f_{e^+}$ & $f_{\bar{p}}$ & $\phi_{e^+}^\odot$ & $\phi_{\bar{p}}^\odot$ & $\langle\sigma v\rangle_{BF}$ & $\langle\sigma v\rangle_{\rm CMB}$ & $\chi^2_{\rm min}(e^+)$ & $\chi^2_{\rm min} (\bar{p})$ \\\hline
$1000$ & $10$ & $29.6$  & $0.77$  & $0.1198$ & $7.52\times10^{-27}$  & $0.78$ & $1.28$  & $600$ & $1019$ & $7.35\times10^{-24}$ & $3.25\times10^{-24}$ & $91.48$ & $75.75$ \\\hline
$1500$ & $10.5$ & $48.7$  & $0.80$ & $0.1194$ & $1.39\times10^{-26}$  & $0.80$ & $1.28$  & $612$ & $1019$ & $1.42\times10^{-23}$ & $4.90\times10^{-24}$ & $81.28$ & $75.75$ \\\hline
$2000$ & $12$ & $74.7$  & $0.91$  & $0.1191$ & $1.45\times10^{-26}$  & $0.81$ & $1.28$  & $620$ & $1019$ & $2.29\times10^{-23}$ & $6.56\times10^{-24}$ & $110.61$ & $75.75$ \\\hline
\end{tabular}
\caption{Similar with table~\ref{tab:benchzz}, but for $NN\to Z^\prime H_0$ annihilation channel. Here we set $g^\prime=3\times10^{-3}$ which is compatible with the LHC $Z\to4\mu$ direct search limit for light $Z^\prime$ boson. The values of $M_{H_0}$ have been chosen such that the resulted Sommerfeld enhancement factors are match to fitted boost factors, i.e., $S_E\simeq BF$.}
\label{tab:benchzh}
\end{table*}

\subsection{Constraints from other Indirect Detections}
\label{sec:constraints}

\subsubsection{AMS-02 antiproton constraint}

\begin{figure}
\begin{center}
\includegraphics[width=0.45\linewidth]{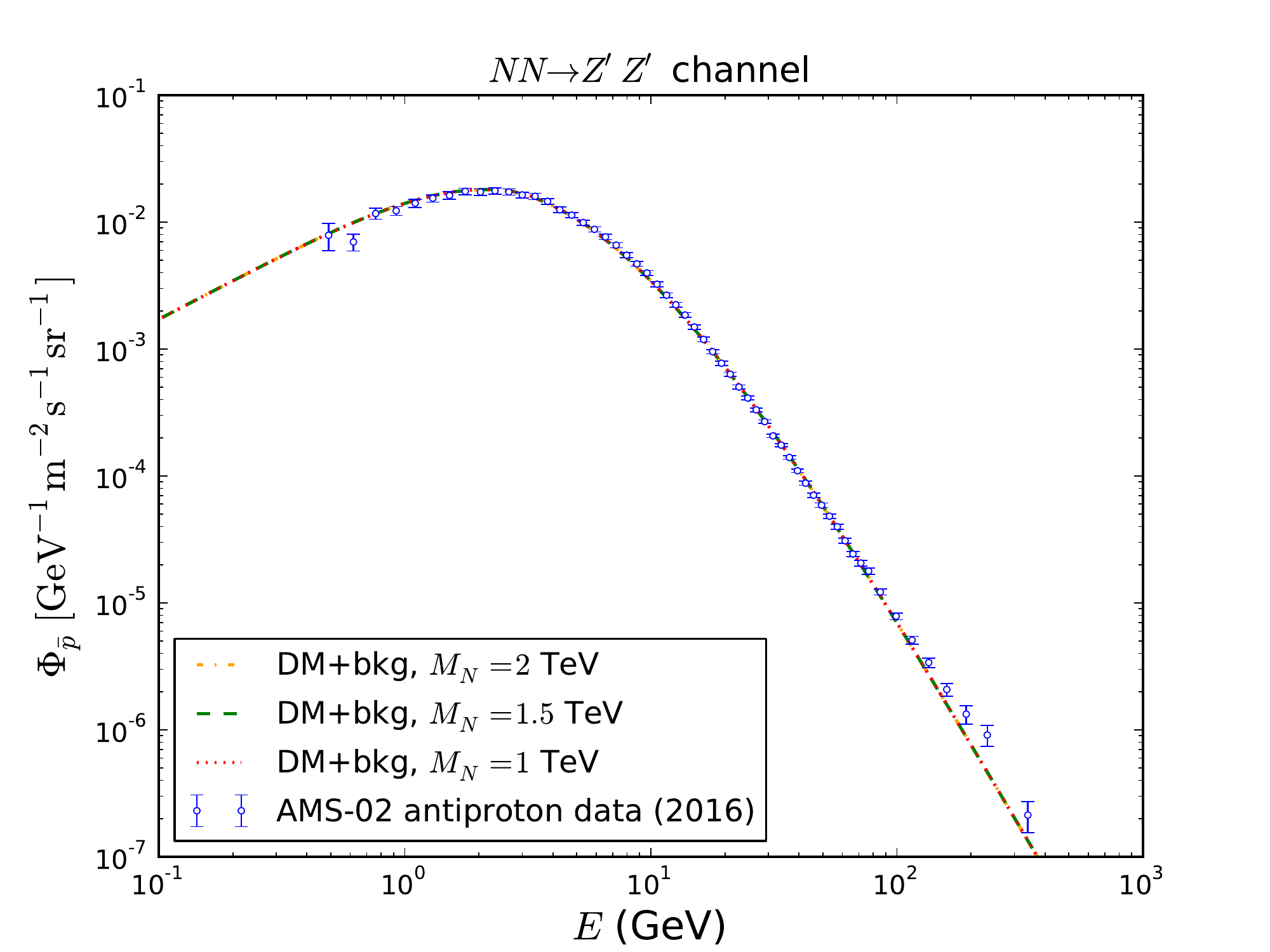}
\includegraphics[width=0.45\linewidth]{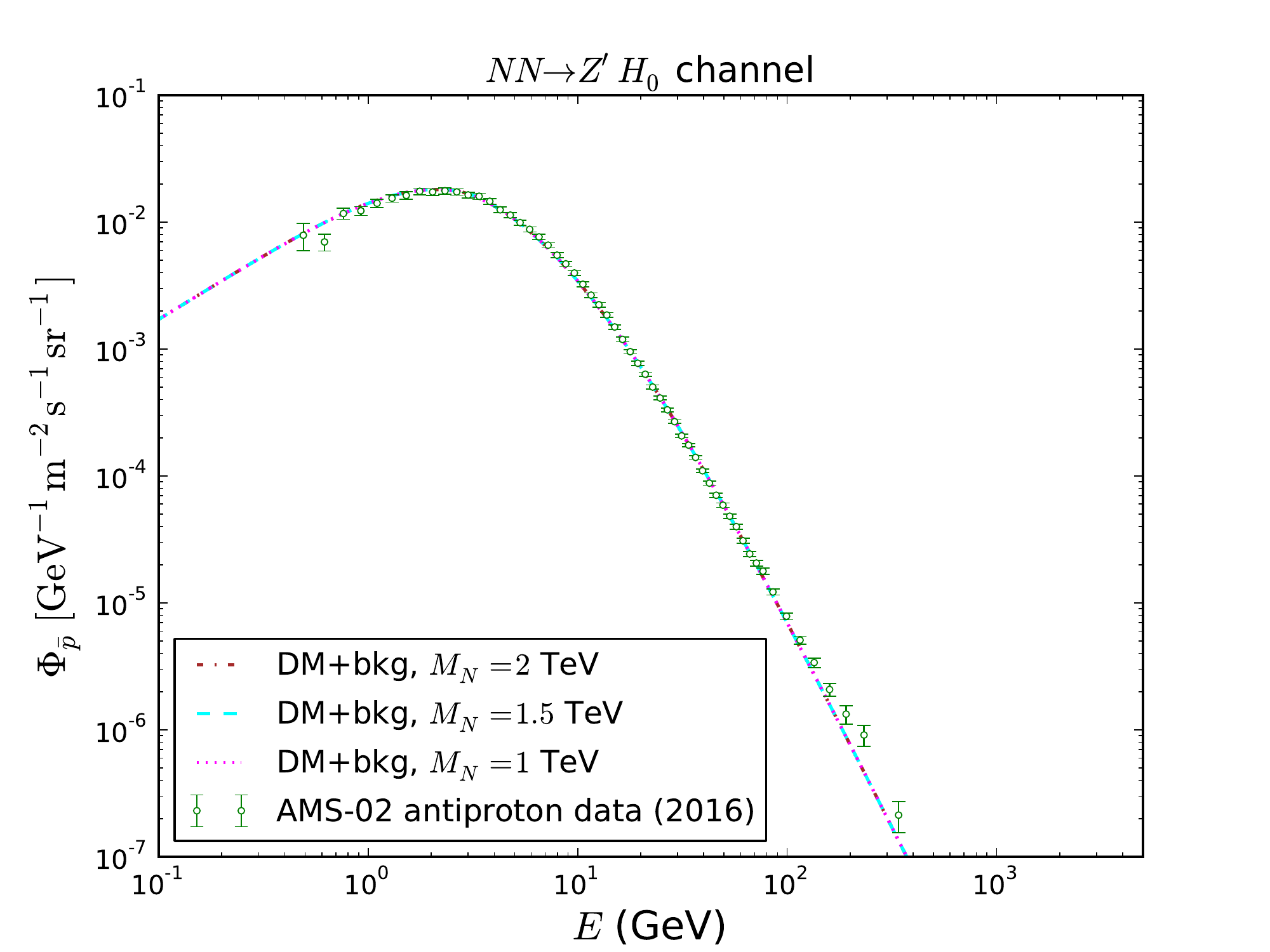}
\end{center}
\caption{The antiproton flux predicted by benchmarks in table~\ref{tab:benchzz} and~\ref{tab:benchzh} with the AMS-02 data.}
\label{Fig:ap}
\end{figure}

Given the benchmarks to explain the positron excess, we now exam the constraints from other indirect detections. As we mentioned in section~\ref{sec:po}, the relevant limits come from antiproton flux, EGRB measurement and impact of energy deposition on CMB anisotropy. From Eq.~(\ref{eq:Yukawa-eff}), the effective coupling of $Z^\prime \bar{d}_i d_j$ leads to antiproton flux. Although highly suppressed by Yukawa-like matrices $\bm{L}^d_{ij}$ and $\bm{R}^d_{ij}$, we still need to investigate whether the predict antiproton flux conflicts with current observation. Similar with Eqs.~(\ref{eq:frac}) and~(\ref{eq:chipo}), the antiproton flux and $\chi^2$ function are respectively given by
\begin{eqnarray}
\Phi_{\bar{p}}(f_{\bar{p}},\phi_{\bar{p}}^\odot, BF) &=& f_{\bar{p}} \Phi_{\bar{p}}^{\rm bkg}(\phi_{\bar{p}}^\odot)+\Phi_{\bar{p}}^{\rm DM}(\phi_{\bar{p}}^\odot, BF),\\\nonumber
\chi^2_{\bar{p}}(f_{\bar{p}},\Phi_{\bar{p}}^\odot, BF)&=&\sum_i\frac{\left[\Phi_{\bar{p},i}(f_{\bar{p}},\Phi_{\bar{p}}^\odot, BF)-\Phi_{\bar{p},i}^{\rm AMS}\right]^2}{(\sigma_{\bar{p},i}^{\rm AMS})^2}.
\end{eqnarray}
$f_{\bar{p}}$ is normalization factors for antiproton background and also set to vary in the range $[0,~5]$. Note that $f_{\bar{p}},~\phi_{\bar{p}}^\odot$ should be different to $f_{e^+},~\phi_{e^+}^\odot$ since the astrophysical sources and propagation processes are distinct for positron and antiproton. On the other hand, $BF$ should be the same due to the fact that enhancement of annihilation cross section is universal for lepton and quark final states. The resulted antiproton flux for benchmarks to interpret positron excess are plotted in figure~\ref{Fig:ap} with AMS-02 measurement~\cite{Aguilar:2016kjl}, and best fit parameter values and $\chi^2_{\rm min}$ are listed in table~\ref{tab:benchzz} and~\ref{tab:benchzh}. All of benchmarks just have same values for parameters and $\chi^2_{\rm min}$, which means that antiproton flux from DM contribution is so small that $\chi^2$ is entirely determined by background. We thus conclude that our model is totally safe for antiproton constraint.

\subsubsection{Fermi-LAT EGRB constraint}

The next important constraint we consider comes from the EGRB measured by Fermi-LAT collaboration~\cite{Ackermann:2014usa}. For calculation of EGRB flux, we follow the procedures in Ref.~\cite{Kawasaki:2009nr,Yuan:2009xq}. The flux of  at redshift $z$ is given as
\begin{align}
\frac{d\Phi_{\rm EGB}}{dE_\gamma}=\frac{c(1+z)^2\bar{\rho}_{\rm DM}^2BF\langle\sigma v\rangle_0}{8\pi M^2_N}\int_z^\infty dz'\frac{(1+z')^3B(z',m_{\rm min})}{H(z')}\frac{dN}{dE_\gamma'}e^{-\tau(z,z',E_\gamma')}\;,
\label{eq:EGB}
\end{align}
where $E_\gamma'=E_\gamma(1+z')/(1+z)$, $\bar{\rho}_{\rm DM}=\rho_c\Omega_{\rm DM}$ is the average density of DM with $\rho_c$ the critical density of the Universe at present, $H(z)=H_0\sqrt{(\Omega_{\rm DM}+\Omega_b)(1+z)^3+\Omega_\Lambda}$ is the Hubble function.

\begin{figure}
\begin{center}
\includegraphics[width=0.5\linewidth]{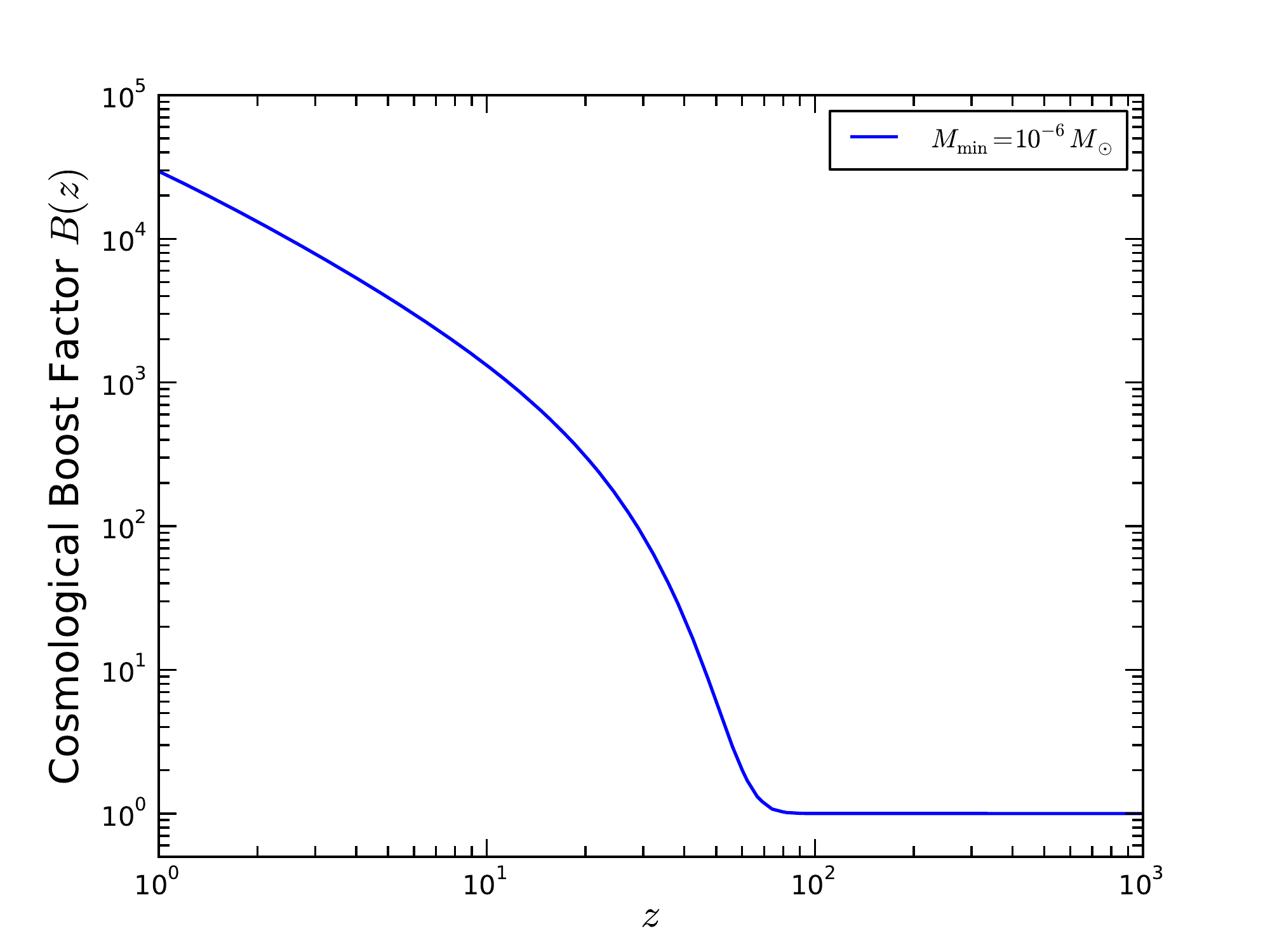}
\end{center}
\caption{Cosmological boost factor $B(z)$ as a function of redshift for Maccio concentration model with $M_{\rm min}=10^{-6}~M_{\bigodot}$.}
\label{fig:bz}
\end{figure}

The $\gamma$-ray generation spectrum for per DM annihilation, $dN/dE_\gamma'$, is dominated by two components:
\begin{align}
\frac{dN}{dE_\gamma'}=\left.\frac{dN}{dE_\gamma'}\right|_{\rm FSR}+\left.\frac{dN}{dE_\gamma'}\right|_{\rm IC}\,,
\end{align}
where $\left.\frac{dN}{dE_\gamma'}\right|_{\rm FSR}$ corresponding to $\gamma$-rays produced from the final state radiation (FSR) of primary charged lepton final states ($\mu$ and $tau$ in our model) due to DM annihilation. $\left.\frac{dN}{dE_\gamma'}\right|_{\rm IC}$ characterizes the $\gamma$-rays resulted from Inverse Compton (IC) scattering between secondary electrons/positrons and CMB photons. For the FSR photon spectrum, we adopt analytical formulas in Refs.~\cite{Bergstrom:2004cy} and~\cite{Fornengo:2004kj}
\begin{eqnarray}
\left.\frac{dN}{dx}\right|^\mu_{\rm FSR}&=&\frac{\alpha_{\rm e.m.}}{\pi}\frac{1+(1-x)^2}{x}\ln\left(s(1-x)/m^2_\mu\right)\,,\nonumber\\
\left.\frac{dN}{dx}\right|^\tau_{\rm FSR}&=&x^{-1.31}(6.94x-4.93x^2-0.51x^3)e^{-4.53x}\,,
\end{eqnarray}
where $\alpha_{\rm e.m.}$ is the fine-structure constant, $s=4M^2_N$ and $x=E/M_N$. The $\gamma$-ray photons from IC component is given as~\cite{Profumo:2009uf}
\begin{align}
\left.\frac{dN}{dE}\right|_{\rm IC}=\int d\epsilon n_\gamma(\epsilon) \int dE_e\frac{dn}{dE_e}\sigma_{\rm KN}(\epsilon,E_e,E)\,.
\end{align}
In above equation, $n_\gamma(\epsilon)$ is the photon number density of background radiation, and $\sigma_{\rm KN}(\epsilon,E_e,E)$ is the Klein-Nishina cross section. $dn/dE_e$ is the electron/positron energy spectrum after propagating, which is related to production spectrum $dN_e/dE'_e$ by following equation
\begin{align}
\frac{dn}{dE_e}=\frac{1}{b(E_e,z)}\int^{m_N}_{E_e}dE'_e\frac{dN_e}{dE'_e}\,,
\end{align}
with $b(E_e,z)\simeq 2.67\times10^{-17}(1+z)^4)(E_e/{\rm GeV})^2{\rm GeV}{\rm s}^{-1}$ is the energy loss rate~\cite{Profumo:2009uf}. Eq.~(\ref{eq:EGB}) contains a cosmological boost factor which account for the effect of DM halo clustering, $B(z)\equiv\langle(1+\delta(z))^2\rangle=1+\langle\delta^2(z)\rangle$. We here adopt a halo model that approximates the matter distribution in the Universe as a superposition of DM halos and
$B(z,m_{\rm min})$ can be expressed as
\begin{align}
B(z,M_{\rm min})=1+\frac{\Delta_c}{3\bar{\rho}_{m,0}}\int^\infty_{M_{\rm min}}dM M \frac{dn}{dM}(M,z)f[c(M,z)].
\end{align}
where $\bar{\rho}_{m,0}$ is the matter density at present, $\Delta_c\simeq 200$ is the overdensity at which the halos are defined and $M_{\rm min}$ is the minimal halo mass used in integration. $\frac{dn}{dM}(M,z)$ is the halo mass function with the universal form
\begin{align}
\frac{dn}{dM}(M,z)=\frac{\bar{\rho}_{m,0}}{M^2}\nu f(\nu)\frac{d\log\nu}{d\log M}.
\end{align}
In above equation, the parameter $\nu=[\delta_c(z)/\sigma(M)]^2$. $\delta_c(z)$ is the critical overdensity and the $\sigma(M)$ is the variance of the linear density field in spheres containing a mean mass $M$. $c(M,z)$ represents the halo concentration parameter function and the function $f(c)$ for the halos with the NFW density profile is given as
\begin{align}
f(c)=\frac{c^3}{3}\left[1-\frac{1}{(1+c)^3}\right]\left[\log(1+c)-\frac{c}{1+c}\right]^{-2}.
\end{align}
We choose Maccio concentration model~\cite{Maccio:2008pcd} and set $M_{\rm min}=10^{-6}~M_{\odot}$ in calculation, and evolution of $B(z,m_{\rm min})$ with redshift is presented in figure~\ref{fig:bz}. In last, $\tau(E_\gamma', z,z',)$ is the the optical depth of $\gamma$-ray photon with energy $E'$ and propagating from $z'$ to $z$. Which can be expressed as
\begin{align}
\tau(E_\gamma',z,z')=\int^{z'}_{z}dz''\frac{\alpha(E_\gamma'',z'')}{H(z'')(1+z'')},
\end{align}
where $\alpha(E_\gamma,z)$ is the absorption coefficient and $E_\gamma''=E_\gamma'(1+z'')/(1+z')$. For detailed description of interactions and corresponding absorption coefficients for photon propagation which are taken into account, see Ref.~\cite{Kawasaki:2009nr,Yuan:2009xq}.

\begin{figure}
\begin{center}
\includegraphics[width=0.45\linewidth]{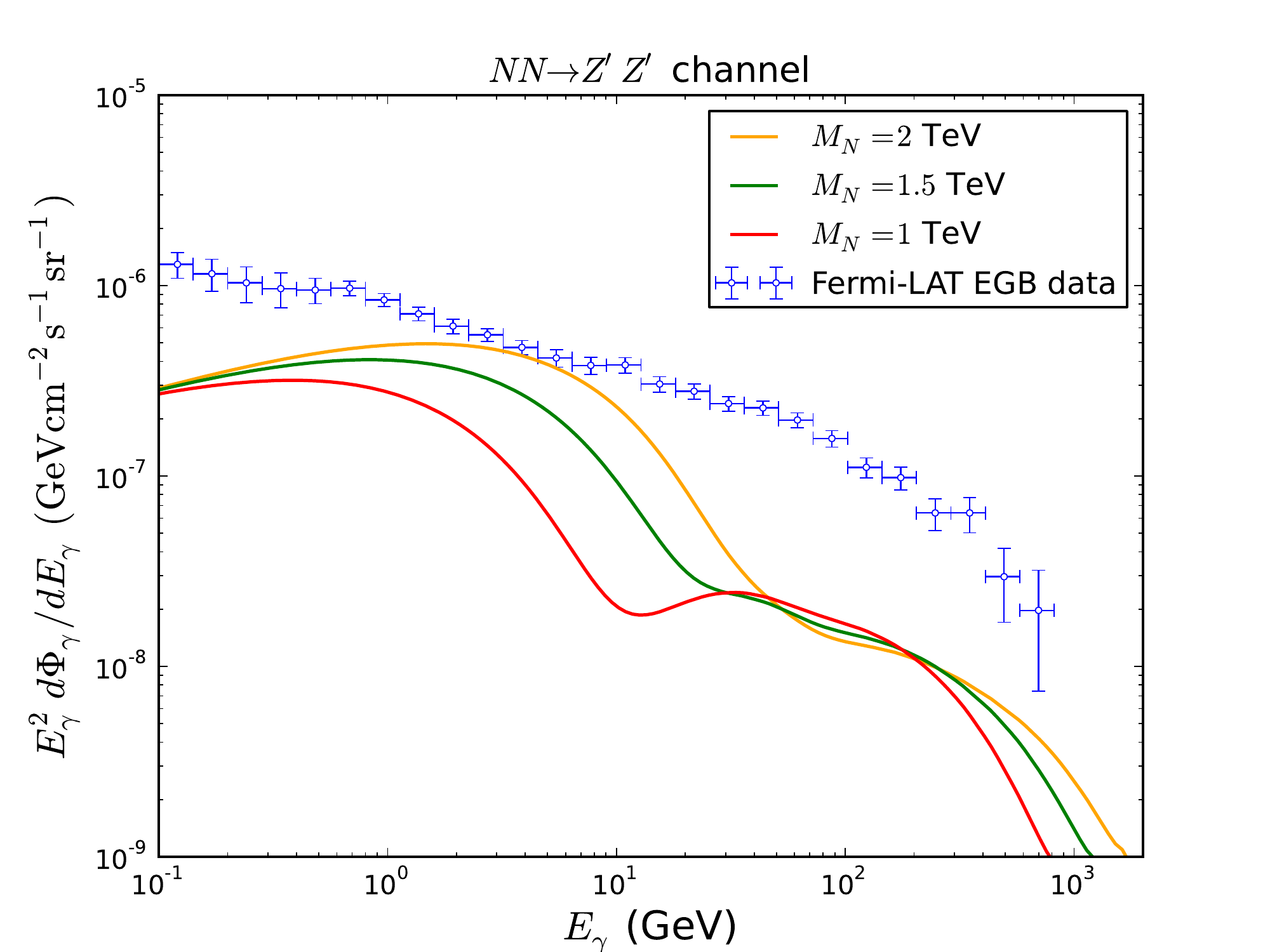}
\includegraphics[width=0.45\linewidth]{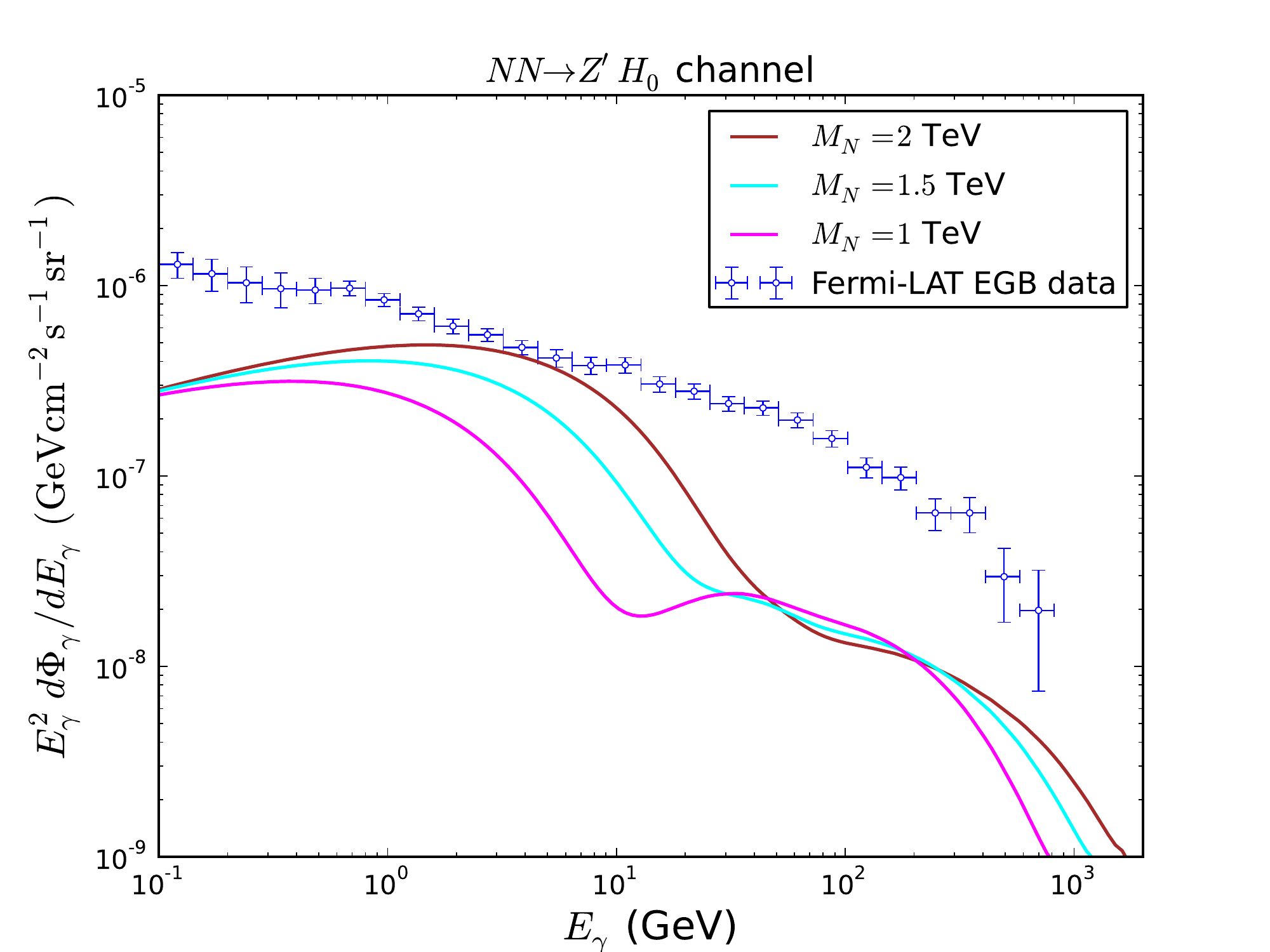}
\end{center}
\caption{Comparison of the EGRB flux produced by benchmarks in tables~\ref{tab:benchzz} and~\ref{tab:benchzh} with the Fermi-LAT measurements.}
\label{fig:EGRB}
\end{figure}

In figure~\ref{fig:EGRB}, we show the total EGRB flux for benchmarks in tables~\ref{tab:benchzz} and~\ref{tab:benchzh}, with the Fermi-LAT measurement~\cite{Ackermann:2014usa}. We found that our benchmarks are marginally compatible with observation. However, for different concentration model and
smaller minimal halo, they could be potentially ruled out.

\subsubsection{Planck CMB constraint}

\begin{figure}
\begin{center}
\includegraphics[width=0.45\linewidth]{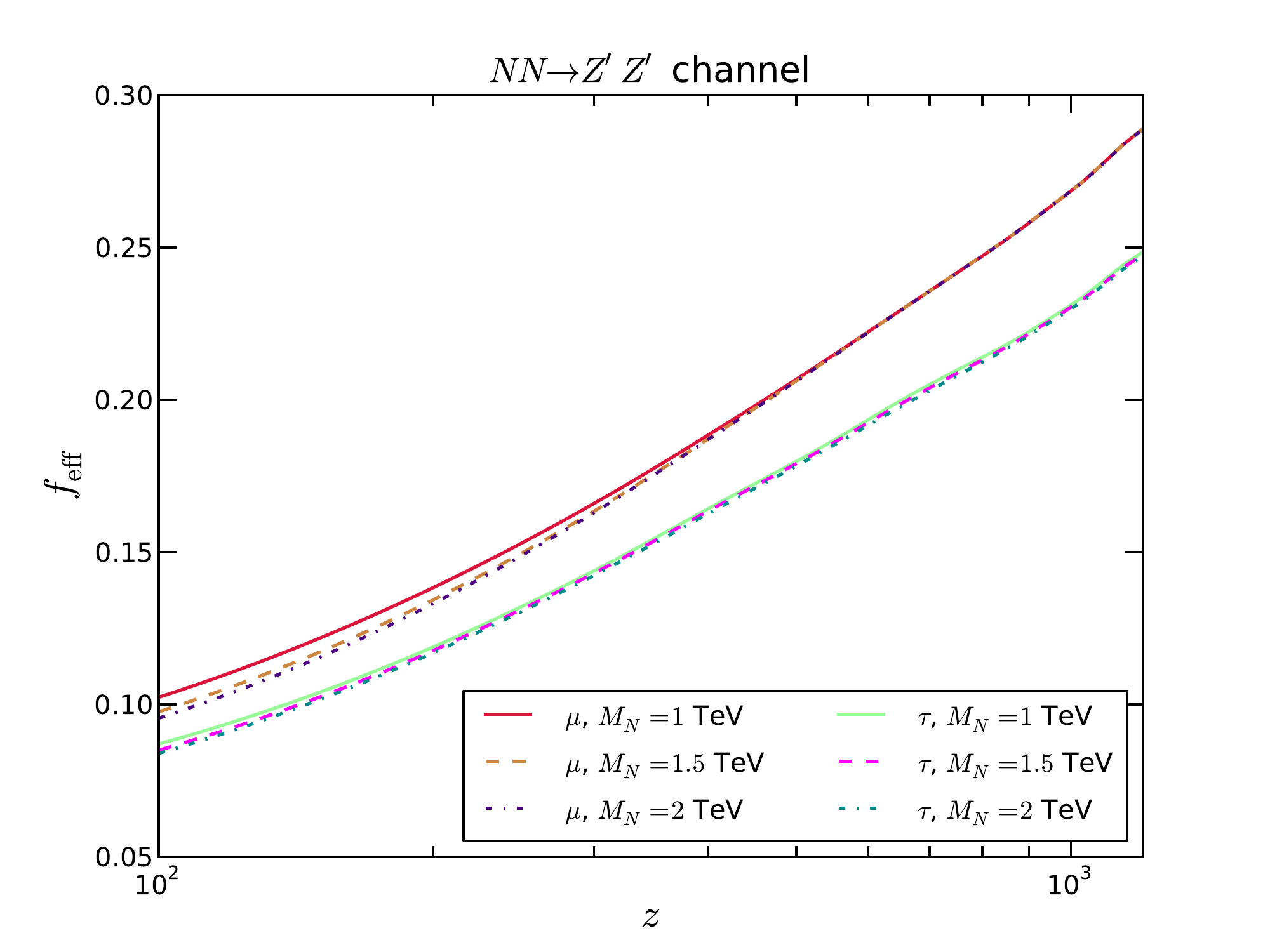}
\includegraphics[width=0.45\linewidth]{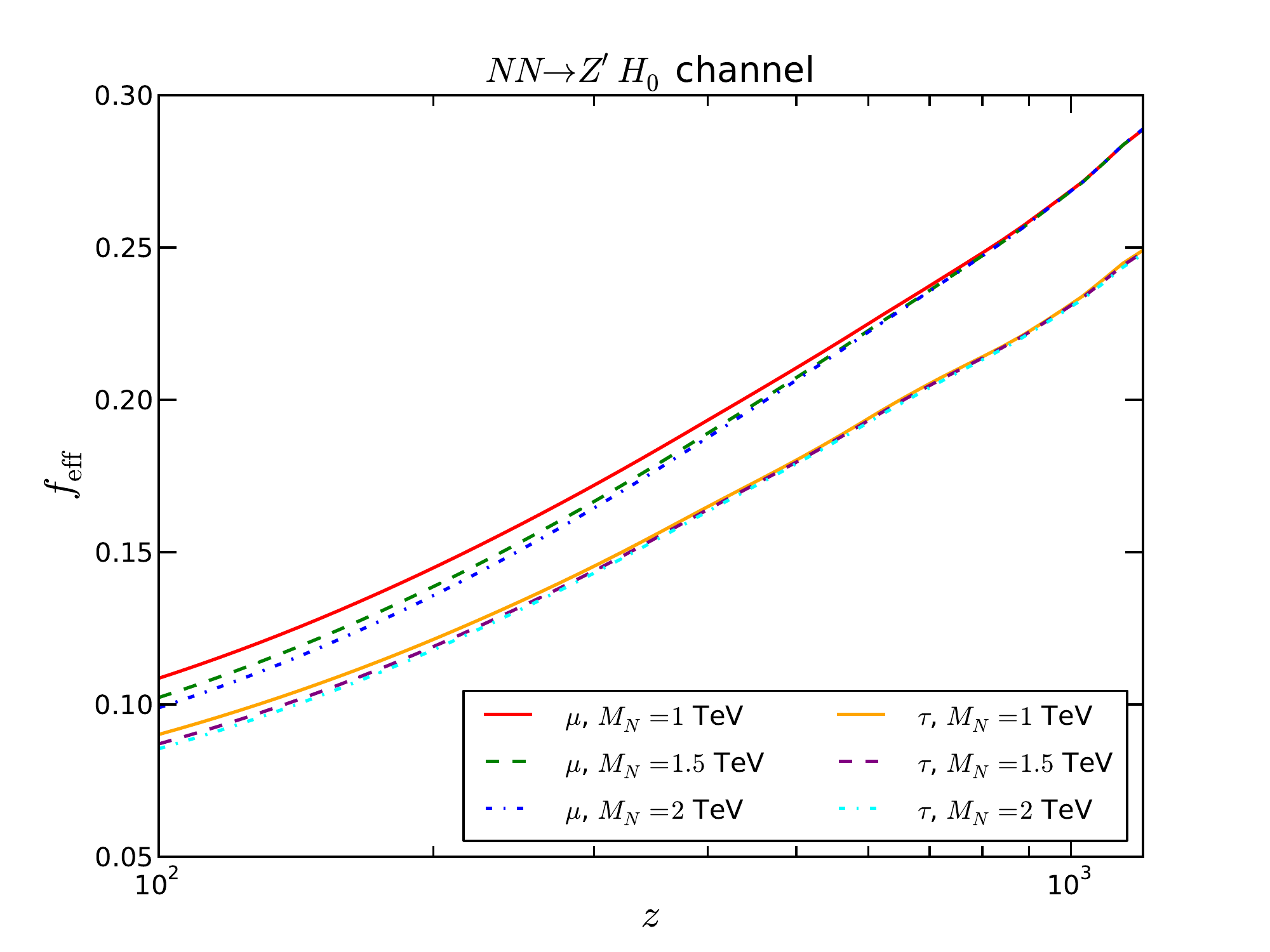}
\end{center}
\caption{$f_{\rm eff}(z)$ curves in the range of $z\in100-1200$ for benchmarks in tables~\ref{tab:benchzz} (left) and~\ref{tab:benchzh} (right).}
\label{fig:eff}
\end{figure}

The last constraint necessarily need to consider is the effect of DM annihilation on CMB anisotropy. Annihilation of DM
to SM particles between recombination and reionization epoch can inject and deposit energy into intergalactic medium (IGM) through produced electrons, positrons and photons via photoionization, Coulomb scattering, Compton processes, bremsstrahlung and recombination. The primary effect of these processes are to alter ionization fraction and left an imprint on spectrum of CMB anisotropy. The injection power into the IGM per unit volume at redshift $z$ is given by as~\cite{Finkbeiner:2011dx}
\begin{align}
\left(\frac{dE}{dVdt}\right)_{\rm injected}=\rho_{\rm DM,0}^2(1+z)^6\frac{g \langle\sigma v\rangle}{M_{\rm DM}},
\end{align}
where $\rho_{\rm DM,0}$ is the present DM density, and the degeneracy $g=1$ for our Majorana DM $N$. the relationship between deposited energy and injected one can be parameterized as
\begin{align}
\left(\frac{dE}{dVdt}\right)_{\rm deposited}=f_{\rm eff}(z)\left(\frac{dE}{dVdt}\right)_{\rm injected},
\end{align}
where $f_{\rm eff}(z)$ denotes dimensionless efficiency factor. It is conventional to define function $p_{\rm ann}(z)=f_{\rm eff}(z)\langle\sigma v\rangle/M_{\rm DM}$, which contains full information about the CMB constraint. In specific, for a given primary annihilation final state $i$ with the annihilation spectrum of positron $dN^i_{e^+}/dE$ and photon $dN^i_\gamma/dE$, $f_{\rm eff}(z)$ can be weighted as~\cite{Slatyer:2015jla},
\begin{align}
f^i_{\rm eff}(m_{\rm DM},z)=\int^{m_{\rm DM}}_0 dE\frac{E}{m_{\rm DM}}\left[2\frac{dN^i_{e^+}(m_{\rm DM},E)}{dE}f^{e^+e^-}_{\rm eff}(E,z)+\frac{dN^i_\gamma(m_{\rm DM},E)}{dE}f^\gamma_{\rm eff}(E,z)\right].
\label{eq:eff1}
\end{align}
The detailed calculation of $f_{\rm eff}(z)$ has been developed in Refs.~\cite{Galli:2009zc,Slatyer:2009yq,Kanzaki:2009hf,Galli:2011rz,Finkbeiner:2011dx,Slatyer:2012yq,Galli:2013dna,Lopez-Honorez:2013lcm} and the numerical results available at~\cite{epsilon} for all of 28 SM final states based on annihilation spectrum provided by {\tt PPPC4DMID} package \cite{Cirelli:2010xx}. In our model, $N$ annihilate into muons, taus, and neutrinos according to channels $NN\to Z^\prime Z^\prime \to 2\ell^+2\ell^-+2\nu_\ell2\nu_\ell$ and $NN\to Z^\prime H_0(\to Z'Z') \to 3\ell^+3\ell^-+3\nu_\ell3\nu_\ell$, ($\ell=\mu,\tau$). Notice that both $Z^\prime$ and $H_0$ are on-shell mediators in relative annihilation channels. Corresponding $f_{\rm eff}(z)$ for our benchmarks can be obtained by applying simple kinematics, and expressed in terms of Eq.~(\ref{eq:eff1}) as follows
\begin{eqnarray}
f^{Z'Z'}_{\rm eff} (z) &=& \sum_{i=\mu,\tau} f^i_{\rm eff}(M_N/2,z), \\ \nonumber
f^{Z'H_0}_{\rm eff}(z) &=& \sum_{i=\mu,\tau}\left[\frac{f^i_{\rm eff}(E_{Z'}/2,z)+2(E_{H_0}/E_{Z'})f^i_{\rm eff}(E_{H_0}/4)}{1+2(E_{H_0}/E_{Z'})}\right],
\end{eqnarray}
respectively for two annihilation channels. In above equation, $E_{Z',H_0}=M_N[1\pm (M^2_{Z'}-M^2_{H_0})/4M^2_N]$. $f_{\rm eff}(z)$ curves for benchmarks in tables~\ref{tab:benchzz} and~\ref{tab:benchzh} are displayed in figure~\ref{fig:eff} for redshift $z\in100-1200$. Here we only care about $\mu,~\tau$ final states since neutrinos have negligible energy deposition. In the later case,  neutrino detectors such as IceCube can impose a better constraint, but current limits are still weak, thus do not threaten our model.

Based on these curves, We now calculate CMB limits on annihilation cross section. As is shown in previous studies by using the method of principal component analysis, the CMB constraint is dominated by the behavior of $f_{\rm eff}(z)$ at $z\sim 600$. This then impose upper limit on $p_{\rm ann}$ function as~\cite{Aad:2015uza},
\begin{align}
p_{\rm ann}=f_{\rm eff}(z=600)\frac{\BR_{\mu,\tau}\langle\sigma v\rangle_{\rm CMB}}{M_N}<3.4\times10^{-28}~{\rm cm}^3/{\rm s}/\GeV (95\%~{\rm C.L.}),
\end{align}
where $\BR_{\mu,\tau}\simeq1/3$ with neglecting phase space difference. Resulting upper bound for annihilation cross section, $\langle\sigma v\rangle_{\rm CMB}$, are also listed in tables~\ref{tab:benchzz} and~\ref{tab:benchzh}. Comparing with $\langle\sigma v\rangle_{BF}$, we found that they have slight conflict. However, considering current measurement allowed a larger local density within uncertainty, which will result in a smaller cross section (by a factor of several times) required by positron excess. In this case, our benchmarks are still comparable with CMB constraint marginally.

\section{Conclusion}\label{sec:CL}

In light of recent $R_{K^{(*)}}$ anomaly and AMS-02 positron excess, we revise the gauged $U(1)_{L_\mu-L_\tau}$ scotogenic model. This model implement gauged $U(1)_{L_\mu-L_\tau}$ symmetry in the one-loop radiative neutrino mass model, where three right-handed neutrinos $N_{\ell}(\ell=e,\mu,\tau)$, a scalar doublet $\eta$, and a scalar singlet $S$ with charge $+1$ under $U(1)_{L_\mu-L_\tau}$ are introduced. The $U(1)_{L_\mu-L_\tau}$ symmetry is broken spontaneously by the VEV of $S$, resulting the massive gauge boson $Z'$. As a complementarity for previous consideration of light $Z'$, we instead mainly consider the case of heavy $Z'$, i.e., $M_{Z'}\gtrsim10~\GeV$.

Provided the existence of certain vector-like quarks charged under $U(1)_{L_\mu-L_\tau}$, an effective $Z' bs$ coupling could be generated, thus the gauge boson $Z'$ will contribute to the process $b\to s \mu^+\mu^-$. In the scenario of heavy $Z'$, the required Wilson coefficient $C_9^\mu\approx-0.95$ to explain $R_{K^{(*)}}$ anomaly can be acquired with Yukawa coupling $Y_Q=0.122$ and $M_{Q}=10~\TeV$. Meanwhile, constraints on $Z'$ mainly come from neutrino trident production and $B_s$ mixing, which actually  require $550~\GeV\lesssim M_{Z'}/g'\lesssim 4~\TeV$. For $Z'\lesssim 50~\GeV$, the search for $Z'$ in the $Z\to 4\mu$ final states set a more tight constraint than neutrino trident production.

As for fermion DM $N$, the gauge boson $Z'$ and scalar singlet $H_0$ arising from the gauged $U(1)_{L_\mu-L_\tau}$ provide viable annihilation channels. To illustrate the effects of various annihilation processes, we implement a random scan over certain parameter space under constraints from relic density and direct detection. For $M_{Z'}>M_N$, the $NN\to Z^{\prime *}\to \ell^+\ell^-,\nu_\ell\nu_\ell(\ell=\mu,\tau)$ are the dominant channel, while $NN\to Z'Z', Z'H_0,H_0H_0$ channels become dominant for $M_{Z',H_0}<M_N$. Especially, the $NN\to Z'Z', Z'H_0(\to Z'Z')$ channels can account for observed positron excess.

Finally, our combined analysis with previous two part shows that the $R_{K^{(*)}}$ anomaly and AMS-02 positron excess can be explained simultaneously under constraints from neutrino trident production, $B_s$ mixing, neutrino mixing, DM relic density, direct detection as well as various indirect detection (AMS-02 antiproton, Fermi-LAT EGRB and CMB measurements).

\section*{Acknowledgements}
RD thanks Qiang Yuan for help on EGRB calculation, also thanks Tracy R. Slatyer and Chen Sun for useful discussion and help on CMB constraint. This work was supported by the National Natural Science Foundation of China under Grant No. 11805081, Natural Science Foundation of Shandong Province under Grant No. ZR2019QA021.

%%%%%%%%%%%%%%%%%%%%%%%%%%%%%

\end{document}